\documentclass[twocolumn]{aastex631}
\usepackage{amsmath,amssymb}
\usepackage{graphicx}
\usepackage{natbib}
\usepackage{hyperref} 
\usepackage[nameinlink]{cleveref} 
\usepackage{nameref}
\usepackage{CJK}
\usepackage{pifont}
\usepackage{tablefootnote}

\usepackage{placeins}
\usepackage{afterpage}

\newcommand{\ms}{m~s$^{-1}$\;}

\begin{document}
\begin{CJK*}{UTF8}{gbsn}

\title{Breaking the Mass Inclination Degeneracy of Radial Velocity Measurements via Monitoring von Zeipel-Lidov-Kozai Cycles: Implications in the HD 41004 System}

\author[0009-0001-3213-1406]{Zhizhen Qin (秦至臻)}
\affiliation{School of Physics and Astronomy, Sun Yat-sen University, Zhuhai 519082, China}
\affiliation{CSST Science Center for the Guangdong-HongKong-Macau Great Bay Area, Sun Yat-sen University, Zhuhai 519082, China}

\author[0000-0002-9442-137X]{Shang-Fei Liu (刘尚飞）}
\affiliation{School of Physics and Astronomy, Sun Yat-sen University, Zhuhai 519082, China}
\affiliation{CSST Science Center for the Guangdong-HongKong-Macau Great Bay Area, Sun Yat-sen University, Zhuhai 519082, China}
\email{liushangfei@mail.sysu.edu.cn}

\author[0000-0002-0378-2023]{Bo Ma (马波)}
\affiliation{School of Physics and Astronomy, Sun Yat-sen University, Zhuhai 519082, China}
\affiliation{CSST Science Center for the Guangdong-HongKong-Macau Great Bay Area, Sun Yat-sen University, Zhuhai 519082, China}

\author[0000-0001-6039-0555]{Fabo Feng (冯发波)}
\affiliation{Tsung-Dao Lee Institute, Shanghai Jiao Tong University, Shengrong Road 520, Shanghai 201210, China}
\affiliation{School of Physics and Astronomy, Shanghai Jiao Tong University, 800 Dongchuan Road, Shanghai 200240, China}

\begin{abstract}
We investigate the dynamical stability of the S-type planet in the compact binary HD 41004. Using $N$-body simulations, we find that the planet could be dynamically stable at a mutual angle inclination up to $\sim75^\circ$. The von Zeipel-Lidov-Kozai (vZLK) mechanism becomes active when the mutual inclination is greater than 39.2$^\circ$. High-inclination orbits exhibit coupled oscillations in eccentricity and inclination, along with apsidal precession. Synthetic radial velocity (RV) modeling shows that these secular variations produce measurable signatures across a broad range of timescales, from full vZLK cycles to observationally accessible decades. For instance, a high mutual inclination at 75$^\circ$ can induce RV drifts exceeding 5 \ms per planetary orbit ($\sim 1.9 \,\text{m s}^{-1}\, \text{yr}^{-1}$) in circular binary configurations. The presence of eccentric vZLK further accelerates these drifts, enhancing the detectability. Long-term RV observations of this system offer a unique pathway to dynamically constrain the orbital inclination and thus determine the true mass of HD 41004 Ab. The degeneracy of mass inclination is well known when using RV measurements alone. 
Our results highlight that HD 41004 Ab and potentially other S-type planets in compact binaries are promising targets for breaking such a degeneracy by studying the dynamics induced by the vZLK mechanism through long-term high-precision RV monitoring.
\end{abstract}

\section{Introduction} 
\label{sec:intro}

Although a significant fraction of stars are expected to reside in binary systems \citep{1991Duquennoy&Mayor}, the number of planets discovered in such systems remains relatively small. This disparity reflects both the observational limitations and the dynamical evolution inherent in binary environments \citep{Musielak2005, Su_2021, Stevenson_2023}. As of March 2025, just over 200 planets have been identified in more than 150 binary systems\footnote{\url{https://adg.univie.ac.at/schwarz/multiple.html}}. More than half of these exoplanets are detected through radial velocity (RV) surveys \citep{Su_2021}, complemented by transit photometry and high-contrast direct imaging.

Planetary orbits in binaries are generally classified as either S-type, where the planet orbits one of the two stars, or P-type, where the planet orbits the binary pair \citep{Dvorak2002PlanetsIH}. The majority of confirmed planets in such systems are S-type, including well-studied cases like $\gamma$~Cephei Ab \citep{1988Campbell, Hatzes_2003}. 

However, RV method detects periodic velocity changes of a star along the observer's line of sight, i.e. the amplitude of perturbations is proportional to the minimum mass of a planet. The true mass of the planet and its orbital inclination (relative to the sky plane) are degenerate parameters via the RV measurement alone, which are crucial to characterize hierarchical systems. Recent efforts have aimed to overcome these limitations. For instance, \citet{Stevenson_2023} proposed improved strategies for detecting S-type planets, while \citet{2023Chauvin} employed a combination of RV measurements, speckle interferometry, high-contrast imaging, and astrometry to revisit the dynamically complex HD~196885 system. In addition, \citet{huang2025closebyhabitableexoplanetsurvey} utilized dynamical RV fitting to better constrain the orbital parameters of S-type planets in close binaries.

The dynamical architectures of binary planetary systems offer critical insights into planet formation and long-term evolution \citep{1978Heppenheimer,Haghighipour_2006,Xie2010,Gong_2018,Fragione_2018,Camargo2023}. Notably, \citet{Su_2021} finds that S-type planets in binaries typically exhibit higher orbital eccentricities than those around single stars. Close binaries, in particular, appear to host a distinct population of massive, short-period planets, suggesting unique dynamical histories \citep{Zucker_2002,Su_2021}.

The complex dynamical architectures of binary systems have motivated extensive theoretical and numerical studies on planetary stability and habitability. Pioneering work by \citet{Holman_1999} establishes an empirical expression for the critical semi-major axis beyond which planetary orbits become unstable. Subsequent studies employ different methods from direct integration to chaos indicators such as the Fast Lyapunov Indicator (FLI; \citealt{lyapinov, Froeschl1997THEFL}) and MEGNO \citep{megno} to explore the role of mutual inclination, eccentricity, and orbital spacing in shaping long-term stability \citep{2002Pilat-Lohinger,Dvorak2002PlanetsIH, Hill_binary,2003Pilat-Lohinger,Satyal_2013,2014Satyal,2020Quarles,2022Troy,2024Bhaskar}. For instance, \citet{Dvorak2002PlanetsIH} shows that moderate inclinations ($\lesssim 50^\circ$) can enlarge the stability region, while \citet{Hill_binary} and \citet{Satyal_2013} systematically mapped the stability boundaries using Hill criteria \citep{Hill, Mako_2004} and chaos diagnostics. The stability of inclined orbits in binary-planet and multi-planet systems has also been extensively investigated \citep{2003Pilat-Lohinger, 2020Quarles, 2022Troy, 2004Veras, 2024Bhaskar}.

The stability results discussed above highlight the critical role of mutual inclination in shaping the dynamical evolution of planetary systems in compact binaries. Beyond determining orbital stability, mutual inclination also governs the onset of long-term secular interactions. In hierarchical triple systems, the dominant mechanism responsible for such interactions is the von Zeipel-Lidov-Kozai (vZLK) mechanism \citep{vonZeipel,LIDOV1962719,1962Kozai}. This mechanism drives long-term oscillations in eccentricity and inclination, accompanied by apsidal and nodal precession, under the influence of a distant, inclined companion. Originally developed to describe asteroid dynamics perturbed by Jupiter, the vZLK mechanism is most effective for mutual inclinations between $39.2^{\circ}$ and $140.8^{\circ}$, and has since been extended to a wide range of astrophysical contexts \citep{Harrington1968,Wu_2003,Ford_2004,Fabrycky2007, 2013Teyssandier,Naoz_2013,2019Ito}. In compact binaries such as $\gamma$~Cephei \citep{heintz_1982,1999Pourbaix,Pourbaix_2016}, the eccentric vZLK extension has been used to account for the influence of nonzero eccentricities in both the inner and outer orbits \citep{Lithwick_2011, Li_2014, Naoz_2016, Huang_2022}. The same mechanism may have reshaped the stability landscape of some other compact binary systems, such as HD~196885 and $\alpha$~Centauri \citep{2014Satyal,Quarles_2016,Giuppone2017}.

However, most of these studies focus on the stability of vZLK dynamics. Thus, it is natural to ask whether such secular effects could manifest in real planetary systems—and under what conditions they become observable. Among compact binaries, HD~41004 stands out as a particularly compelling case for studying vZLK-induced dynamics due to its tight configuration and massive stellar companion. Discovered via the pioneering application of multi-order TODCOR to echelle spectra \citep{Zucker1994, Santos_2002, Zucker_2003}, HD~41004 is a hierarchical four-body system consisting of a primary K0V star (HD~41004 A), its planetary companion (HD~41004 Ab), and a close binary companion (HD~41004 B + Bb), composed of an M2V star and a brown dwarf (see Table~\ref{tab:HD 41004_table}). The planet orbits at 1.7 AU, while the secondary binary lies at a projected distance of $\sim$22 AU, enabling strong secular perturbations. The system's complexity has prompted advanced modeling efforts; for instance, \citet{Andrade-Ines2016} emphasizes the need for second-order perturbation theory \citep{Laskar2010}, and \citet{2016Satyal} identifies a critical inclination near $65^{\circ}$ for dynamical stability. Despite its dynamical richness, previous observations place only weak constraints on the system parameters, and the long-term secular evolution of HD~41004—particularly in the high mutual inclination regime—remains poorly understood.  

If the mutual inclination between the planetary and binary orbits exceeds the vZLK threshold, the system may exhibit long-term orbital variations, including eccentricity and inclination oscillation, and secular precession — potentially detectable via RV monitoring, transit timing variations, or high-precision astrometry.

In this study, we investigate the dynamical stability of the HD~41004 system and explore the long-term observational signatures in radial velocity associated with vZLK cycles. By combining theoretical modeling, numerical simulations, and dynamical analysis, we aim to clarify the conditions under which vZLK effects are triggered and to identify potential observational diagnostics. The theoretical framework, governing equations, and simulation setup are detailed in Section~\ref{sec:methods}. Simulation results are presented in Section~\ref{sec:results}, followed by a discussion in Section~\ref{sec:discussion}, and a summary of key findings in Section~\ref{sec:conclusion}.

\begin{figure*}[ht]
\centering
\includegraphics[width=0.95\columnwidth]{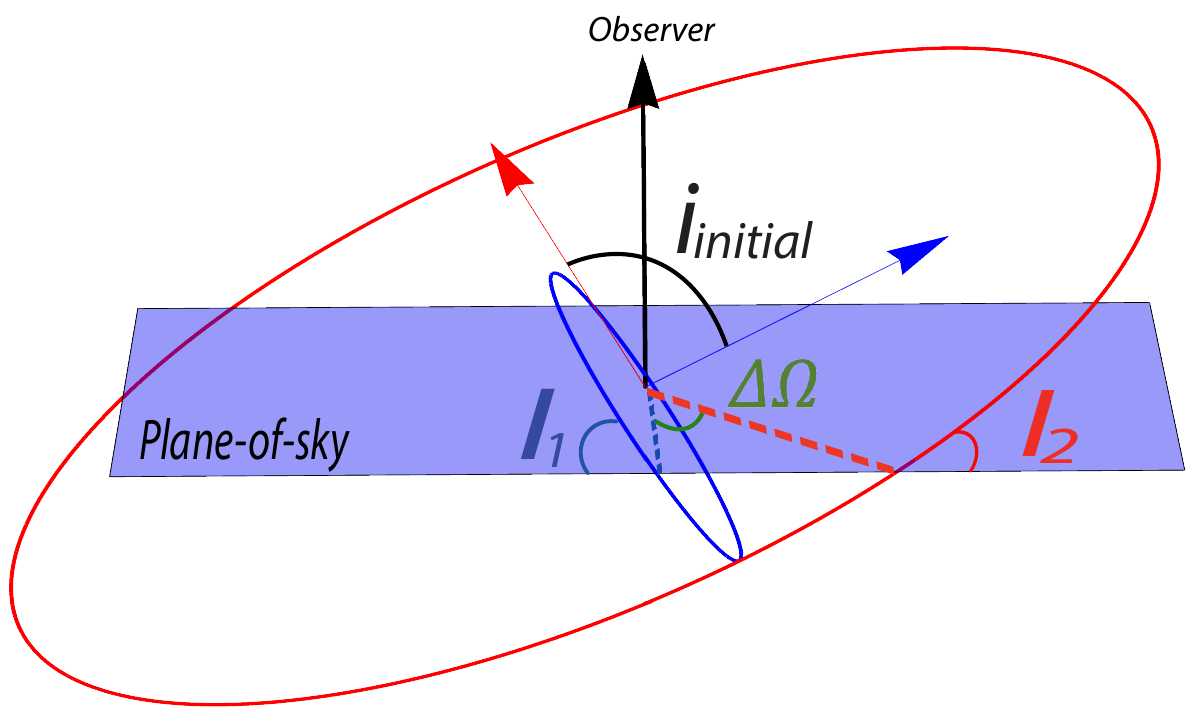}
\includegraphics[width=0.95\columnwidth]{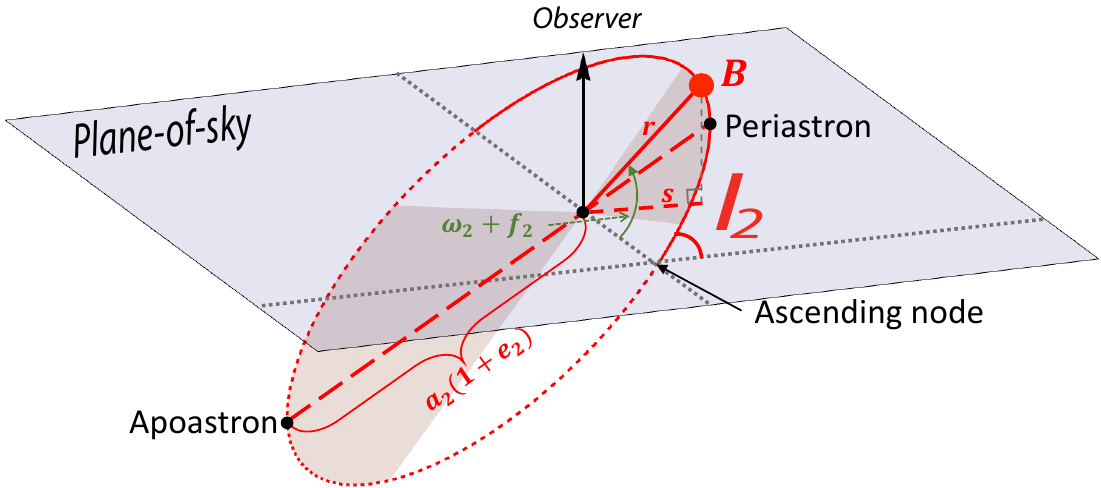}
\caption{
\textbf{Schematic geometry of a hierarchical three-body system and projected separation $s$ of HD~41004~B.} 
Left panel: the blue plane represents the plane of the sky, perpendicular to the observer's line of sight (black arrow). Red and blue arrows denote the orbital angular momentum vectors of the outer and inner orbits, respectively. The angles $I_1$ and $I_2$ are their inclinations relative to the plane of the sky, while $i_{\text{initial}}$ is the mutual inclination between the two orbits. The green angle $\Delta \Omega$ represents the difference in the longitudes of ascending nodes. Although $I_1$ and $I_2$ depend on the observer's orientation, both $i_{\text{initial}}$ and $\Delta \Omega$ are intrinsic dynamical parameters of the system. Right panel: the red ellipse shows a representative orbit of component B with eccentricity $e_2$ and inclination $I_2$. The gray sectors illustrate orbital phase angle $\omega_2 + f_2$ within $\pm60^\circ$ around periastron and apoastron. The projected separation $s$, instantaneous distance $r$, and inferred semi-major axis $a_2$ follow the definitions in Equation~\ref{eq:a_2}.}
\label{fig:inclination_schematic}
\end{figure*}

\section{Methods} \label{sec:methods}
\subsection{Orbital Configuration and Parameter Setup of HD 41004}
\label{sec:triple_config}

In hierarchical triple systems, the orbital configuration can be described by two nested orbits: an inner orbit between a central body $m_0$ and an inner companion $m_1$, and an outer orbit between the center of mass of the inner pair and an external perturber $m_2$. Each orbit is characterized by its semi-major axis, eccentricity, inclination relative to a chosen reference plane (commonly the plane of the sky), and orientation angles (longitude of ascending node $\Omega$ and argument of periastron $\omega$).

The mutual inclination $i_{\text{initial}}$ between the inner and outer orbits is a key parameter governing secular interactions, including the vZLK mechanism. This inclination is defined as the angle between the orbital angular momentum vectors of the inner and outer orbits and is related to the observable orbital inclinations $I_1$, $I_2$ and the nodal difference $\Delta \Omega$ via:
\begin{align}
\label{eq:inc}
\cos{i_\text{initial}} = \cos{I_1}\cos{I_2}+\sin{I_1}\sin{I_2}\cos{\Delta\Omega}.
\end{align}
While $I_1$ and $I_2$ are orientation-dependent quantities tied to the observer's line of sight, $i_{\text{initial}}$ is an intrinsic property of the system's dynamical configuration and remains invariant under changes in viewing geometry (see the left panel of Figure \ref{fig:inclination_schematic}).

In the context of radial velocity (RV) observations, the true masses of orbiting companions are not directly measured; instead, only the minimum masses $m_1 \sin I_1$ and $m_2 \sin I_2$ are inferred. These are given by \citep{Lee_2003}:

\begin{align}
\label{RV_equation_general}
K_1 &= \left( \frac{2\pi \textrm{G}}{P_1} \right)^{1/3} \frac{m_1 \sin I_1}{(m_0 + m_1)^{2/3}} \frac{1}{\sqrt{1 - e_1^2}}; \notag\\
K_2 &= \left( \frac{2\pi \textrm{G}}{P_2} \right)^{1/3} \frac{m_2 \sin I_2}{(m_0 + m_1 + m_2)^{2/3}} \frac{1}{\sqrt{1 - e_2^2}}.
\end{align}

This mass-inclination degeneracy is critical for the dynamical evolution of the system, as different combinations of inclination, true mass, and eccentricity can produce the same RV signals but lead to different orbital paths and dynamical outcomes.

\renewcommand{\tabcolsep}{5pt}
\begin{deluxetable*}{p{4.2cm}p{4.6cm}p{4.6cm}p{4.6cm}}
\tablecaption{Parameters of the HD~41004 system constrained by observations and explored in our dynamical modeling \label{tab:HD 41004_table}}
\tablehead{
\multicolumn{1}{l}{\textbf{Parameter}} & 
\multicolumn{1}{l}{\textbf{Observation}\textsuperscript{\scriptsize a}} & 
\multicolumn{1}{l}{\textbf{Stability Simulations}} & 
\multicolumn{1}{l}{\textbf{RV Simulations}}
}
\startdata
\multicolumn{4}{l}{\textbf{HD~41004 A}} \\
Mass ($m_0$) & $0.7\,M_{\odot}$ & $0.7\,M_{\odot}$ & $0.7\,M_{\odot}$ \\
\hline
\multicolumn{4}{l}{\textbf{HD~41004 Ab}} \\
Minimum mass ($m_1\sin{I_1}$) & $(2.54\pm0.74)\,M_\text{J}$ & $2.54\,M_\text{J}$ & $1.8, 2.54, 3.28\,M_\text{J}$ \\
RV Semi-amplitude ($K_1$) & $99\pm60 \textrm{m s}^{-1}$ & --- & --- \\
Semi-major Axis ($a_1$) & $1.7\pm0.04$ AU & 1.7 AU & 1.7 AU \\
Eccentricity ($e_1$) & $0.74\pm0.2$ & 0.74 & 0.74 \\
Argument of Pericenter ($\omega_1$) & $97^{\circ}\pm31^{\circ}$ & $97^{\circ}$ & $97^{\circ}$ \\
Line-of-sight Inclination ($I_1$) & --- & determined via Eq.~\eqref{eq:inc} & determined via Eq.~\eqref{eq:inc} \\
Mutual Inclination ($i_\text{initial}$) & --- & $[0^{\circ},90^{\circ}]$ & $0^{\circ},30^{\circ},45^{\circ},60^{\circ},75^{\circ}$\\
\hline
\multicolumn{4}{l}{\textbf{HD~41004 B+Bb}} \\
Mass ($m_2$) & $0.40 + 0.02\,M_{\odot}$ & $0.40 + 0.02\,M_{\odot}$ & $0.40 + 0.02\,M_{\odot}$ \\
Semi-major Axis ($a_2$) & $\sim$22 AU & $[12.2,220.0]$\textsuperscript{\scriptsize b} AU   & 22,25,30,40 AU \\
Eccentricity ($e_2$) & --- & $0, 0.4, 0.8$ & $0, 0.4$ \\
Line-of-sight Inclination ($I_2$) & --- & $0^{\circ}, 45^{\circ}, 90^{\circ}$ & $90^{\circ}$ \\
Nodal Difference ($\Delta\Omega$) & --- & $[0^{\circ},360^{\circ}]$\textsuperscript{\scriptsize c} & $[0^{\circ},360^{\circ}]$\textsuperscript{\scriptsize c}\\
\enddata
\tablecomments{ Square brackets [ ] denote variable parameter ranges, while comma-separated values indicate fixed values in the simulations.
Unspecified orbital parameters are assumed to be zero.\\ \textsuperscript{\scriptsize a} Observation parameters from \citet{Santos_2002,Zucker_2003,Zucker_2004}. $a_2$ denotes the projected binary separation (21-23~AU), not the true semi-major axis.
\textsuperscript{\scriptsize b}
The specific parameter combinations for $a_2$ in the stability simulations are listed in Table 2. 
\textsuperscript{\scriptsize c}
In the simulations, the initial nodal difference $\Delta\Omega = \Omega_1 - \Omega_2$ is set to $0^\circ$ and evolves dynamically over the range $[0^\circ, 360^\circ]$.
}
\end{deluxetable*}

%HD 41004 is a visual binary consisting of a primary K1V star HD 41004 A and a secondary M2V star HD 41004 B with a V-band magnitude difference of 3.7 and an angular separation of $0.5''$ \citep{1997ESA}.
%We adopt $m_0 = 0.7\,M_\odot$ and $m_2 = 0.40+0.02\,M_\odot$. 
%HD 41004 B+Bb is a close binary comprising an M2V star and a brown dwarf \citep{Santos_2002}. 
Following \citet{2017Hamers}, we treat the quadruple HD 41004 system as a hierarchical three-body configuration, approximating HD41004 B+Bb as a single component B, to facilitate the analysis of the long-term orbital evolution of planet Ab. $m_0$, $m_1$, and $m_2$ thus denote masses of the primary star HD 41004 A, the S-type planetary companion HD 41004 Ab, and the compact companion HD 41004 B+Bb, respectively. The complete list of observational and simulation parameters adopted in this study is summarized in Table \ref{tab:HD 41004_table}.

The observational inclination ($I_1$) of planet Ab is derived from the binary inclination ($I_2$) and the mutual inclination ($i_{\text{initial}}$) using Equation~\ref{eq:inc}. In our stability simulations, we adopt a minimum mass of $m_1 \sin I_1 = 2.54\,M_\text{J}$ reported by \citet{Zucker_2004}.

%The mass of HD 41004 B is determined by photometric analysis using Stromgren $uvby$ indices \citep{Santos_2002}.
The orbit of HD~41004~B remains poorly constrained. The sky-projected separation ($s$) of the stellar binary is approximately 22 AU \citep{Zucker_2003,Zucker_2004}. However, the true semi-major axis ($a_2$) may differ substantially from $s$, depending on the inclination ($I_2$), orbital phase angle ($\omega_2 + f_2$), and eccentricity ($e_2$) (see the right panel of Figure \ref{fig:inclination_schematic}). The value of $a_2$ can be estimated using Kepler's laws \citep{2015Greco}:
\begin{align}
\label{eq:a_2}
a_2 &= \frac{r}{1 - e_2^2}(1 + e_2 \cos f_2)\\ \notag
&= \frac{s}{(1 \pm e_2) \sin \theta} \\ \notag
&= \frac{s}{(1 \pm e_2) \sqrt{1 - \sin^2 I_2 \sin^2(\omega_2 + f_2)}}.
\end{align}
To study the dynamical influences of HD 41004 B, we explore a range of $a_2$ in our stability analysis by varying $I_2$ and $e_2$. We adopt $e_2 = 0, 0.4, 0.8$, reflecting the typical eccentricities of stellar binaries \citep{1991Mayor}. Taking the assumption that component B has an orbital phase angle $\omega_2 + f_2$ within $\pm60^\circ$ around either periastron or apoastron (the gray sectors in the right panel of Figure \ref{fig:inclination_schematic}), $a_2$ spans a broad range between different $I_2$ (see Equation~\ref{eq:a_2}) except for the trivial configuration in which $e_2=0$ , $I_2 = 0^\circ$ and thus $a_2 \equiv 22 \;\textrm{AU}$, as summarized in Table~\ref{tab:a2_ranges}. Such a choice covers $\sim67-91\%$ of the orbital period for $e_2$ from 0 to 0.8, and includes both the minimum and maximum semi-major axes consistent with the projected separation.

\begin{table}[h!]
    \centering
    \caption{Suite of orbital parameters of the stellar binary HD 41004 B for the stability analysis of HD 41004 Ab.}
    \label{tab:a2_ranges}
    \begin{tabular}{c|c|c}
        \hline
        \textbf{$e_2$} & \textbf{$I_2$ (deg)} & \textbf{$a_2$ Range (AU)} \\
        \hline
        0 & $0^\circ$  & 22.0\\
            & $45^\circ$ & 22.0 -- 31.1 \\
            & $90^\circ$ & 22.0 -- 44.0 \\
        \hline
        0.4 & $0^\circ$  & 15.7 -- 36.7 \\
            & $45^\circ$ & 15.7 -- 51.9 \\
            & $90^\circ$ & 15.7 -- 73.3 \\
        \hline
        0.8 & $0^\circ$  & 12.2 -- 110.0 \\
            & $45^\circ$ & 12.2 -- 155.6 \\
            & $90^\circ$ & 12.2 -- 220.0 \\
        \hline
    \end{tabular}
\end{table}

For RV simulations including dynamical fitting of existing RV data and observational signatures of RV drifts (the third column of Table \ref{tab:HD 41004_table}), we simply fix $I_2$, as we shall show that the impact of $I_2$ is not significant on a relatively short timescale. However, changes in $a_2$ and $e_2$ can substantially alter RV signals on one planetary orbit.

\subsection{\textit{N}-body Dynamical Modeling}
\label{sec:simulation}
To explore the dynamical stability and potential observational signatures in radial velocity measurements of the HD~41004 system, we perform numerical simulations using the \texttt{rebound} \textit{N}-body integrator \citep{rebound, reboundias15}.Our study focuses on three main objectives: (1) mapping the system's stability across a range of orbital configurations; (2) performing unrestricted three-body dynamical modeling to fit the existing RV data; and (3) generating synthetic high-precision RV datasets to investigate the potential observational signatures of vZLK cycles acting on planet Ab.

\subsubsection{Stability Analysis}
\label{sec:megno_stability}

Earlier work (e.g., \citealt{2016Satyal,Giuppone2017}) approximates the HD 41004 system using a three-dimensional elliptic restricted three-body problem (ER3BP) approximation, in which HD 41004~B and its brown dwarf companion are treated as a single perturbing mass, while planet Ab is modeled as a mass-less test particle.

In this study, we model the system with unrestricted three-body dynamics and assess its stability using the Mean Exponential Growth Factor of Nearby Orbits (MEGNO) indicator (see Appendix~\ref{sec:megno}), which can efficiently distinguish between stable, quasi-periodic (with $\bar{Y} \to 2$) and chaotic orbits (where $\bar{Y}$ diverges). This approach allows us to effectively map dynamical regimes and identify instability boundaries within a vast parameter space.

% While this simplification enables long-term stability exploration, it overlooks the mass-inclination degeneracy intrinsic to radial velocity detections. In such cases, the planet's true mass depends on the orbital inclination-a factor that can substantially influence the system's dynamics, particularly in vZLK-active regimes.

Because the planetary mass has the least impact on the dynamical stability except for the plane-of-sky case ($I_1=0^\circ$), in our unrestricted hierarchical three-body model, the minimum mass of the planet $m_1 \sin I_1$ is fixed ignoring its uncertainty, while $I_2$ and $e_2$ are allowed to vary to explore the parameter space of mutual inclinations ($i_\text{initial}$) and binary separations ($a_2$). The planetary inclination $I_1$ is adjusted via Equation~\eqref{eq:inc} to account for the mass-inclination degeneracy \citep{Stevenson_2023}. For simplicity, we set $\Delta\Omega = 0$ and adopt three representative values of $I_2$ ($0^\circ$, $45^\circ$, $90^\circ$). We run simulations on a uniform $50 \times 50$ grid in $a_2$-$i_\text{initial}$ space and each one is integrated over $2 \times 10^6$ planetary orbits. The relevant parameters are summarized in Tables~\ref{tab:HD 41004_table} and \ref{tab:a2_ranges}.

We perform high-precision integrations using the IAS15 integrator and construct MEGNO maps to identify stable and unstable regions. Also, orbits were deemed unstable if the vZLK mechanism induced strong eccentricity growth, causing the pericenter to fall below the Roche limit. The combined use of MEGNO and the Roche limit criterion offers a reliable approach to assess long-term dynamical stability. We show the results in Section~\ref{sec:stability_results}.

\subsubsection{RV Simulations}
\label{sec:rv_retrieval}
On the timescale of the observation of HD 41004 A, planetary effects dominate the RV variations of the primary star. \citet{Zucker_2003, Zucker_2004} derive a minimum mass $m_1\sin{I_1}$ of $2.54 \pm 0.74~\text{M}_J$ with a Keplerian fit ignoring its stellar companion. However, to extract potential vZLK cycles from long-term RV signals, the binary effects must be taken into account.

%The RV trend of HD 41004 A is primarily driven by the planetary signal, with additional contributions from its stellar companion B. As listed in Table~\ref{tab:HD 41004_table}, the observed RV semi-amplitude of HD 41004 Ab is $99 \pm 60~\text{m/s}$, corresponding to a minimum mass of $2.54 \pm 0.74~\text{M}_J$ based on a Keplerian fit \citep{Zucker_2003, Zucker_2004}. However, the limited observational baseline leaves Because key binary parameters ($a_2$, $e_2$) are poorly constrained, we first adopt dynamical modeling to constrain the binary's influence on the RV signal.}

We first simulate the unrestricted three-body system and generate the synthetic RV signals of HD 41004 A, accounting for gravitational perturbations from HD 41004 Ab and B. The simulated RV signals are then compared with the observational data to evaluate the impacts of varying $a_2$, $e_2$, $i_\text{initial}$ and $m_1 \sin I_1$ (Table \ref{tab:HD 41004_table}). For each configuration, we perform a phase search that spans at least a full orbit of companion B to find the best fit. The RV fitting results are presented in Section~\ref{subsec:rv_fit}.

We further synthesize long-term RV variations of HD 41004 A due to its planetary and stellar companions in configurations that are stable and consistent with RV observations. To identify potential vZLK oscillations in RV signatures, we perform a two-step procedure to isolate the planetary component: (1) integrating the full three-body system to obtain the total RV signal of HD 41004 A, and (2) subtracting the modeled binary-induced contribution. The results of long-term RV variations in possible configurations are summarized in Section \ref{sec:rv_results}
%This enabled analysis of the planet's orbital evolution and its RV impact. To assess the effect of the vZLK mechanism, we vary the mutual inclination from $0^\circ$, where it is suppressed, to $75^\circ$, where it becomes significant, accounting for the mass-inclination degeneracy.

%\textbf{The RV fitting results are presented in Section~\ref{subsec:rv_fit}, and the vZLK-induced planetary RV signatures are discussed in Section~\ref{sec:rv_results}.}

\section{Results}
\label{sec:results}
\subsection{Stability of the HD 41004 System}
\label{sec:stability_results}

The empirical stability criterion developed by \citet{Holman_1999} and refined by \citet{Ballantyne_2021} provides a practical baseline for evaluating the stability of S-type planetary orbits in binary systems. The critical semi-major axis $a_{\text{critical}}$, beyond which planetary orbits are expected to become unstable, is given by:
\begin{align}
a_{\text{critical}} =&\ a_2(0.464 - 0.380\mu - 0.631e_2 \notag \\
&\quad + 0.586\mu e_2 + 0.150e_2^2 - 0.198\mu e_2^2),
\end{align}
where $a_2$ is the binary semi-major axis, $e_2$ the binary eccentricity, and $\mu = m_2 / (m_0 + m_2)$ the binary mass ratio.
For the HD~41004 system, adopting $a_2 = 22$~AU and $\mu \approx 0.375$, the critical semi-major axes are $a_{\text{critical}} \approx 7.07$~AU, $3.72$~AU, and $0.90$~AU for $e_2 = 0$, $0.4$, and $0.8$, respectively. The planet HD~41004~Ab, with $a_1 = 1.7$~AU, lies well within the stability threshold for low to moderate binary eccentricities.

However, this criterion may not be legitimate if the mutual inclination is high. Secular perturbations such as the vZLK mechanism are not considered either. \citet{2016Satyal} has identified a stability threshold at a mutual inclination near $65^\circ$ in the HD~41004 system based on the Hill Stability Function. But the analysis is limited by the test-particle approximation adopted for the planet. Besides, parameters in their study also differ from those in \citet{Zucker_2004}, which may affect their stability results.

\begin{figure*}[htbp]
\centering
\includegraphics[width=2.0\columnwidth]{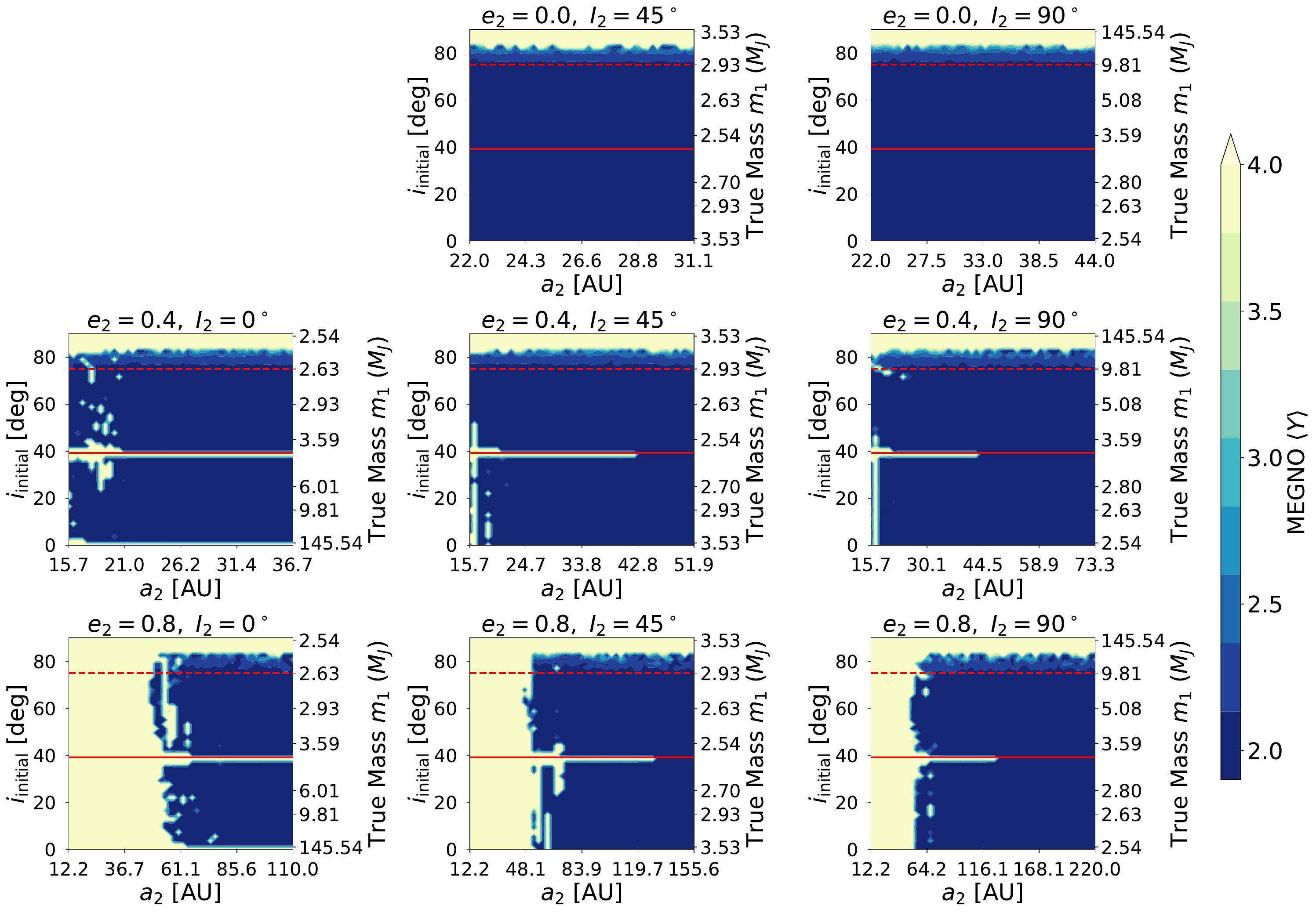}
\caption{
\textbf{MEGNO stability maps for the HD~41004 system under varying binary eccentricities and observational inclinations.}
Each panel depicts the dynamical stability of planet Ab for specific combinations of binary eccentricity ($e_2$) and inclination ($I_2$). The $x$-axis represents the binary semi-major axis $a_2$, and the left $y$-axis shows the mutual inclination $i_\mathrm{initial}$. The right $y$-axis displays the true mass $m_1$ inferred from $m_1 \sin I_1 = 2.54\,M_\text{J}$. The color shading indicates MEGNO values: blue ($\bar{Y} \simeq 2$) represents stable, quasi-periodic motion, while yellow-white ($\bar{Y} > 2$) denotes chaotic behavior. The red dashed line marks $i_\mathrm{initial} = 75^\circ$, and the solid line represents the critical inclination of $39.2^\circ$ for vZLK excitation. 
For $e_2 = 0$, stability is maintained up to $75^\circ$ across all $a_2$. At $e_2 = 0.4$ and $e_2 = 0.8$, instability emerges at smaller $a_2$ and near $i_\mathrm{initial} = 39.2^\circ$ due to enhanced eccentric vZLK forcing. The top-left panel is blank, as $a_2$ is fixed for $I_2 = 0$ and $e_2 = 0$, making simulations unnecessary.
}
\label{fig:MEGNO}
\end{figure*}

In figure~\ref{fig:MEGNO} we plot the MEGNO stability maps of eight non-trivial configurations (see Table \ref{tab:a2_ranges}. The configuration of $e_2 = 0$ and $I_2 = 0$ is excluded, because the binary orbit is fixed by geometry). The horizontal axis in each panel indicates the binary semi-major axis $a_2$ confined by orbital configurations. The left vertical axis of each panel denotes the initial mutual inclination $i_{\text{initial}}$, which determines the planetary inclination $I_1$ and the true mass $m_1$ (right vertical axis) by Equation \ref{eq:inc} and the observed minimum mass. The solid red line marks the critical inclination of $39.2^\circ$ for the onset of the vZLK oscillations, while the dashed line at $75^\circ$ denotes the approximate instability threshold.

The stability maps suggest that whether a hierarchical system is stable or not is mainly determined by its mutual inclination ($i_{\text{initial}}$) and perturbations from the stellar companion. In the first row of Figure~\ref{fig:MEGNO}, when the binary orbit is circular ($e_2 = 0$), HD 41004 Ab remains stable up to $i_{\text{initial}} \sim 75^\circ$ for the entire range of $a_2$. In the second row, when the binary orbit has a mild eccentricity ($e_2 = 0.4$), the planet becomes unstable due to enhanced gravitational perturbations near the binary's periastron, where the reduced separation $a_2 ( 1- e_2)$ undermines stability. In addition, planetary orbits are unstable at the mutual inclination close to the onset of the vZLK effect, \(i_{\text{initial}} = 39.2^\circ\). This unstable region arises from vZLK-induced eccentricity and inclination oscillations, with \(\omega_1\) circulating (\(0^\circ\)-\(360^\circ\)) rather than librating around \(90^\circ\), thus preventing resonant locking and leading to long-term instability. This is consistent with previous findings \citep{2014Satyal, Giuppone2017}, i.e. high $a_1/a_2$ ratios and mutual inclinations around $39.2^\circ$ as well as extreme mutual inclinations could trigger instability. Note that the planet could be very massive when $I_2 = 0^\circ$ and $i_{\text{initial}} \sim 0$. However, the likelihood is rather remote. Highly eccentric cases ($e_2 = 0.8$) in the bottom row further demonstrate that maintaining long-term stability requires even larger binary separation $a_2$.

To summarize, numerical simulations reveal a persistent stability limit at  $i_{\text{initial}} \sim 75^\circ$, which remains robust across variations in planet Ab's true mass and observational inclination. %Near this boundary, the inferred mass ($m_1$) reaches approximately $10\,M_\mathrm{J}$ for $I_2 = 90^\circ$ and $2.9\,M_\mathrm{J}$ for $I_2 = 45^\circ$. \textbf{ Although the derived true mass has minimal impact on dynamical stability, it may induce significant RV variations over long-term evolution.}

\subsection{Dynamical Fitting of Existing RV data}
\label{subsec:rv_fit}
\begin{figure*}[htbp]
\centering
\includegraphics[width=2\columnwidth]{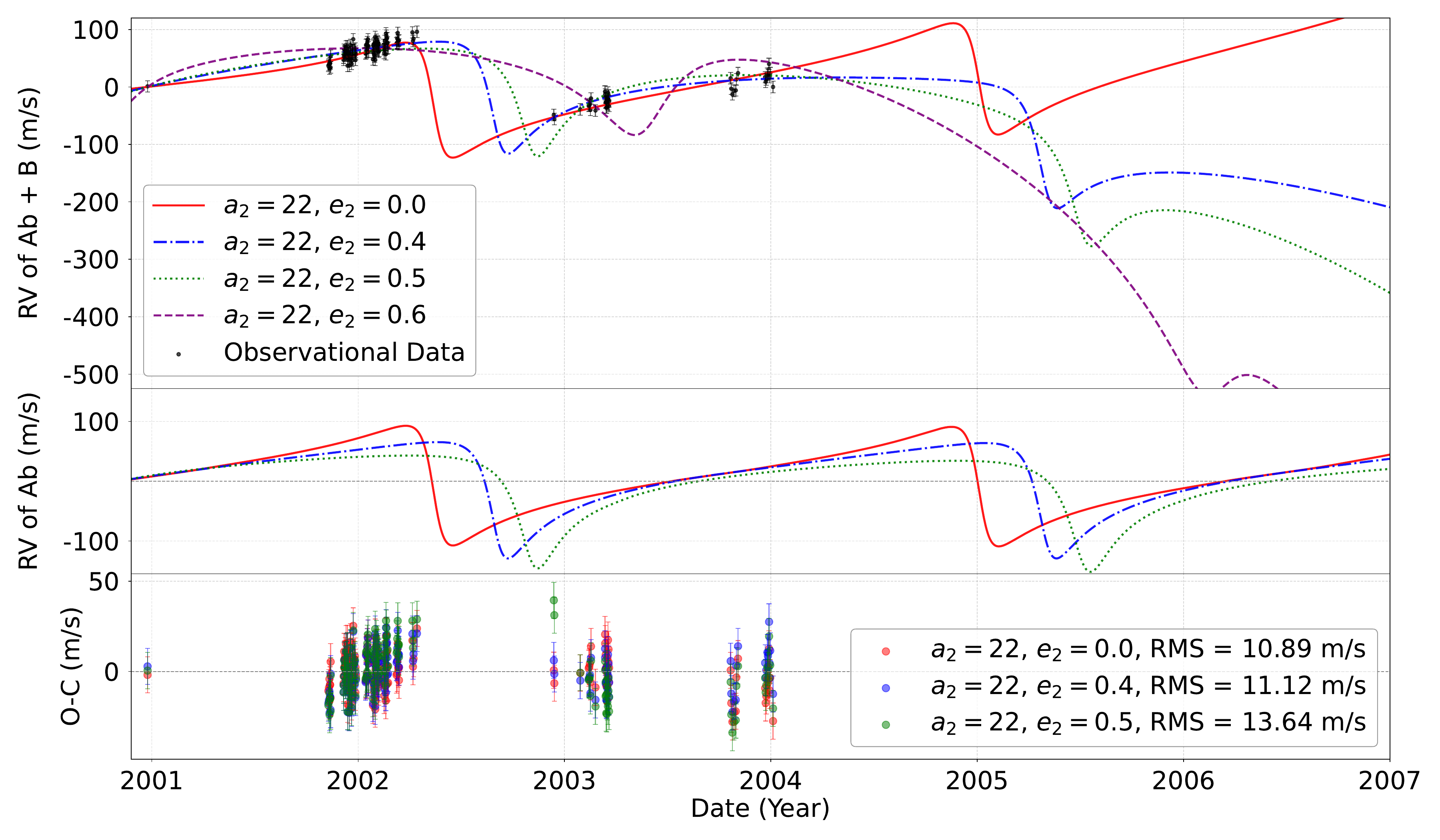}
\includegraphics[width=2\columnwidth]{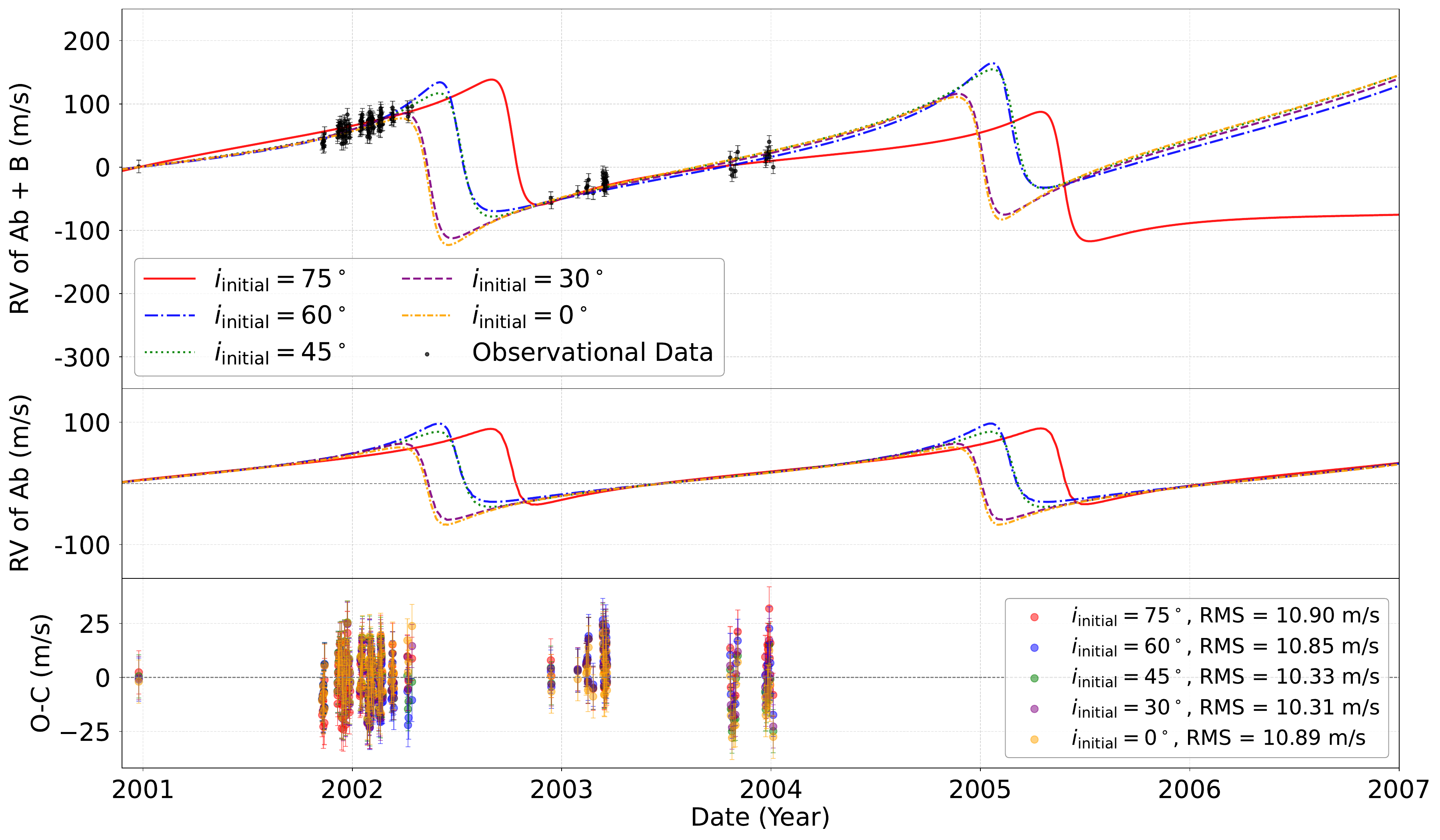}
\caption{
\textbf{Dynamical RV Fitting of HD~41004~A under different orbital configurations.} 
In each plot, the top panel displays the combined RV signal of HD~41004~A, incorporating contributions from both the planetary companion (Ab) and the stellar companion (B). 
The middle panel presents the isolated planetary signal after subtracting the RV contribution of companion B. 
The bottom panel shows the O-C (Observed minus Calculated) residuals for each fit. 
Different colors and line styles represent varying eccentricities of the stellar companion ($e_2$) or different initial mutual inclinations ($i_{\text{initial}}$), depending on the scenario. 
The top figure corresponds to $a_2 = 22$ AU with different $e_2$ and fixed $i_{\text{initial}} = 0^\circ$, while the bottom figure shows $a_2 = 22$ AU with $e_2 = 0$ and varying $i_{\text{initial}}$. 
The legend in each panel provides the RMS values of the O-C residuals for each configuration. 
Observed RV data points are plotted with error bars, and the systemic velocity of $42.513~\textrm{km s}^{-1}$ is subtracted for consistency \citep{Zucker_2004}. 
The time axis is referenced to JD~2451902.77432 (2000 December), marking the epoch of the first published RV measurement.
}
\label{fig:RV_fit}
\end{figure*}

To constrain the orbital architecture of HD 41004, we perform unrestricted three-body simulations to fit the RV observations. Figure~\ref{fig:RV_fit} displays two representative cases with fixed parameters $m_1\sin{I_1} = 2.54\,M_\textrm{J}$ and $a_2 = 22~\text{AU}$: (a) varying $e_2$ at $i_{\text{initial}} = 0^\circ$ (top panel), and (b) varying $i_{\text{initial}}$ at $e_2 = 0$ (bottom panel). In each panel, we plot the combined RV signals of Ab and B, the isolated planetary contribution after binary signal subtraction, and the observed-minus-calculated (O-C) residuals, respectively. Distinct colors and line styles denote various configurations, while data points with error bars are RV observational data, and the systemic velocity of $42.513~\textrm{km s}^{-1}$ is subtracted for consistency \citep{Zucker_2004}. The root-mean-square (RMS) values for each model are indicated in the legends, with the time axis referenced to JD~2451902.77432, the epoch of the first RV measurement.

In the top panel of Figure~\ref{fig:RV_fit}, configurations with $e_2 = 0$ and $e_2 = 0.4$ achieve RMS residuals of $\sim10~\textrm{m s}^{-1}$, consistent with the Keplerian fit reported by \citet{Zucker_2004}. For $e_2 \geq 0.5$, the RMS values increase by $\geq 25\%$, indicating significantly degraded fits. These optimal fits correspond to specific orbital phases of companion B, which induce phase shifts in planet Ab's orbit and drive distinct RV evolutionary patterns across eccentricity regimes.

In the bottom panel of Figure~\ref{fig:RV_fit}, fitting accuracy remains consistent with observations across all $i_{\text{initial}}$ values, with optimal fits corresponding to specific orbital phases of companion B. These phase-dependent configurations induce offsets in planet Ab's RV signal, thereby generating distinct evolutionary patterns in the combined RV curve of component A.

We extend the analysis to broader parameter spaces—including $a_2$ and $m_1 \sin I_1$—with detailed results in Appendix~\ref{sec:rv_fitting_2}. Dynamical fits for wide binaries ($a_2>22$ AU) with eccentricities $e_2\lesssim 0.5$ consistently yield RMS residuals of $\sim10~\textrm{m s}^{-1}$, aligning with the observational precision. While extreme values of the minimum planet mass ($m_1 \sin I_1$)— whether significantly larger or smaller than the nominal $2.54\,M_\textrm{J}$ — systematically degrade the fitting quality, as evidenced by elevated RMS residuals.

In summary, our unrestricted three-body fits constrain the HD 41004 binary to a low-to-moderate eccentricity regime ($e_2\lesssim 0.5$) and yield a minimum planetary mass $m_1 \sin I_1\sim2.54\,M_\textrm{J}$. However, degeneracies prevent tight constraints on the binary separation $a_2$ and initial mutual inclination $i_\mathrm{initial}$. These limitations underscore the necessity of future high-cadence RV observations of this system to resolve its full orbital architecture.

%{Our analysis shows that three-body fits with moderate eccentricities of the binary consistently achieve RMS values around 10~m/s. For wider separations ($a_2 > 22$~AU), this precision is preserved at low $e_2$, while slightly larger values (up to $\sim 0.6$) remain acceptable. In contrast, higher eccentricities generally yield poorer fits, suggesting that the binary likely resides in a low to moderate eccentricity regime. \textcolor{red}{While the current observational baseline is limited, the models do not exclude a close binary orbit near $\sim 22$~AU with low to moderate $e_2$. Future observations will be essential to refine these constraints and determine the true orbital configuration of component B.}}

\subsection{vZLK-induced Planetary RV Signatures}
\label{sec:rv_results}
%Building on the stability analysis in Section~\ref{sec:stability_results}, which demonstrates that HD 41004 Ab can maintain high-inclination orbits,\textbf{ we further investigate its dynamical configuration using the unrestricted three-body model described in Section~\ref{subsec:rv_fit}.  \textcolor{red}{The RV fitting results exhibit consistent accuracies across various system configurations, indicating that a close binary configuration with $a_2 \sim 22$ AU and low to moderate eccentricities ($e_2$) is dynamically plausible. This setup is compatible with both the stability analysis and observed RV variations.}}

Both dynamical stability analysis and RV fitting permit HD 41004 Ab to maintain a high-inclination orbit ($i_\mathrm{initial} \lesssim 75^\circ$ ) relative to the binary plane. When $i_{\text{initial}}$ exceeds the critical angle of $39.2^\circ$, the vZLK mechanism activates. In compact eccentric binaries, the eccentric vZLK mechanism further amplifies and accelerates the excitation of the planet's eccentricity and inclination.

The vZLK timescale governs the coupled oscillations of planetary inclination and eccentricity. In the test-particle approximation, the vZLK timescale follows the analytical expression \citep{Naoz_2016,antognini2015timescales}:
\begin{equation}
\label{eq:Timescale}
t_{\textrm{vZLK}}\simeq \frac{16}{15}\left(\frac{a_{2}^3}{a_{1}^{3/2}}\right)\sqrt{\frac{m_{0}}{\textrm{G}m_{2}^2}}(1 - e_{2}^2)^{3/2},
\end{equation}
which yields $1.02 \times 10^4$ yr at $e_2 = 0$ and $7.89 \times 10^3$ yr at $e_2 = 0.4$ assuming $a_2 = 22$ AU for the HD~41004 system.

Beyond the test-particle approximation, the vZLK timescale exhibits significant parameter dependence \citep{antognini2015timescales, Naoz_2016, 2021Hamers}. We quantify this through numerical simulations across mutual inclinations $i_{\text{initial}} \in \left[40^\circ, 75^\circ\right]$. For $a_2 = 22$ AU: 
\begin{itemize}
    \item At $e_2=0$: $t_\textrm{vZLK}$ decreases from $\sim 9.0 \times 10^3$ yr to $\sim 2.3 \times 10^3$ yr 
    \item At $e_2 = 0.4$: $t_\textrm{vZLK}$ decreases from $\sim 4.5 \times 10^3$ yr to $\sim 1.7 \times 10^3$ yr 
\end{itemize}
with increasing $i_{\text{initial}}$ (see Appendix~\ref{sec:vZLK_timescale} for full parameter scans). Higher mutual inclinations systematically shorten vZLK cycles, accelerating planet Ab's dynamical evolution and enhancing its detectability via time-domain signatures in RV and transit surveys.

%{Since the HD~41004 system lies outside the parameter space explored in \citet{antognini2015timescales}, we conduct numerical simulations to evaluate the vZLK timescales across different mutual inclinations. Detailed results are provided in Appendix~\ref{sec:vZLK_timescale}. Our analysis reveals that higher mutual inclinations ($i_{\text{initial}}$) lead to shorter vZLK cycles, accelerating the dynamical evolution of Ab and enhancing its detectability in observational campaigns.}

Following the RV signal retrieval method (Section~\ref{sec:rv_retrieval}), we isolate the RV signals of HD~41004~Ab across initial mutual inclinations $i_{\text{initial}} \in \left[0^\circ, 75^\circ\right]$. Our simulations span $2 \times 10^6$ days ($\sim$ 5479 yr), sufficient to resolve full vZLK cycles for $i_{\text{initial}} \geq 45^\circ$.

\begin{figure*}[htbp]
\centering
\includegraphics[width=2\columnwidth]{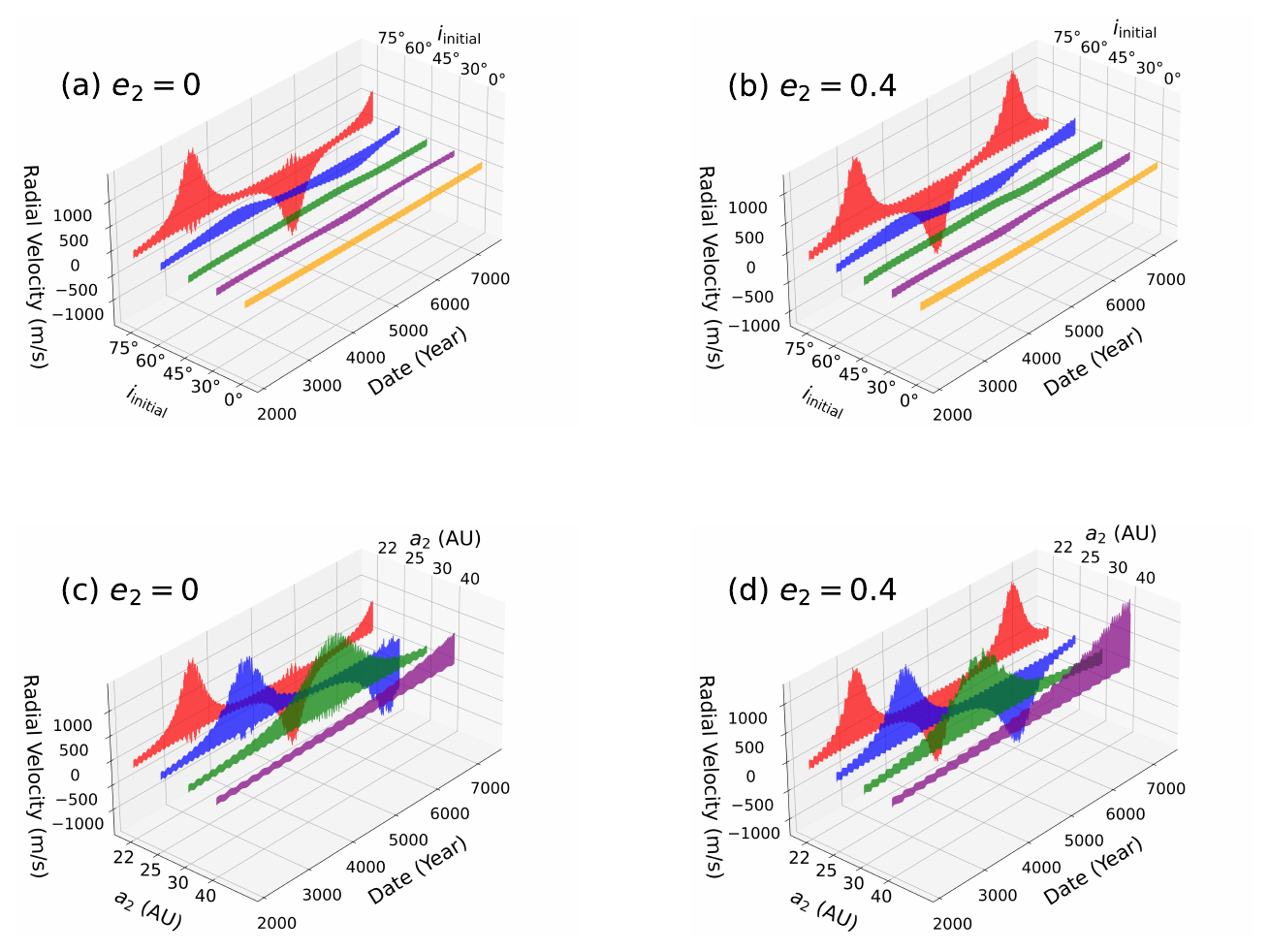}
\caption{
\textbf{Simulated RV curves of HD~41004~Ab for various configurations over $2 \times 10^6$ days ($\sim 5479$ yr).}  
The panel (a) corresponds to $a_2 = 22$ AU, $e_2 = 0$, and $m_1\sin{I_1} = 2.54~M_\textrm{J}$, with varying initial mutual inclinations ($i_{\text{initial}}$). The panel (b) shows the same configuration except with $e_2 = 0.4$. In the panel (c), the simulation explores $a_2 = 22, 25, 30, 40$ AU, $e_2 = 0$, and $m_1\sin{I_1} = 2.54~M_\textrm{J}$, with $i_{\text{initial}} = 75^\circ$. The panel (d) follows the same setup as the bottom-left, but with $e_2 = 0.4$. }
\label{fig:3D}
\end{figure*}

In Figure~\ref{fig:3D} we show the simulated RV curves of HD~41004~Ab with a minimum mass $m_1\sin{I_1} = 2.54~M_\textrm{J}$ under distinct dynamical configurations. In panels (a) and (b) we fix $a_2 = 22$ AU and compare the results of varying initial mutual inclinations $i_{\text{initial}}$ with a binary eccentricity of $e_2 = 0$ and $e_2 = 0.4$, respectively. Both show prominent vZLK-driven RV oscillations at $i_{\text{initial}} \geq 60^\circ$. In panels (c) and (d) we fix $i_{\text{initial}} = 75^\circ$ and compare the results of varying different binary separation $a_2$ with a binary eccentricity of $e_2 = 0$ and $e_2 = 0.4$, respectively. Both highlight enhanced RV variations due to large mutual inclination

More specifically, these signatures result from the combined influence of the following effects:
\begin{enumerate}
    \item The initially high mutual inclination ($i_{\text{initial}}$), which corresponds to a lower observational inclination ($I_1$), leads to a larger true planetary mass. This amplification enhances the RV signal as $I_1$ increases and eccentricity is driven up during vZLK cycles. The maximum semi-amplitude $K_1$ reaches nearly $750~\textrm{m s}^{-1}$ for $i_{\text{initial}} = 75^\circ$ and approximately $150~\textrm{m s}^{-1}$ for $i_{\text{initial}} = 60^\circ$, coinciding with peak $I_1$ and $e_1$. Secular eccentricity oscillations modulate the RV curve shape, especially near periastron. In contrast, for lower mutual inclinations, $K_1$ remains stable around $100~\textrm{m s}^{-1}$.

    \item Apsidal precession, induced by the secular interaction, gradually shifts the orientation of the orbit. This in turn alters the symmetry of the stellar reflex motion, eventually leading to a reversal in the net RV trend—from predominantly positive to negative values.

    \item When $e_2 \neq 0$, eccentric vZLK is activated, shortening the vZLK timescale. As shown in panel (b), the RV variations become more rapid compared to panel (a) across different initial inclinations.

    \item Larger $a_2$ extends the vZLK timescale. In panel (c), the minimum planetary mass remains constant, variations in $K_1$ across different $a_2$ are minimal. However, the slower vZLK evolution leads to more gradual RV modulations. In panel (d), increasing $e_2$ shortens the vZLK timescale, accelerating Ab's orbital evolution, which is reflected in the RV curve.
\end{enumerate}

\begin{figure*}[htbp]
\centering
\includegraphics[width=2\columnwidth]{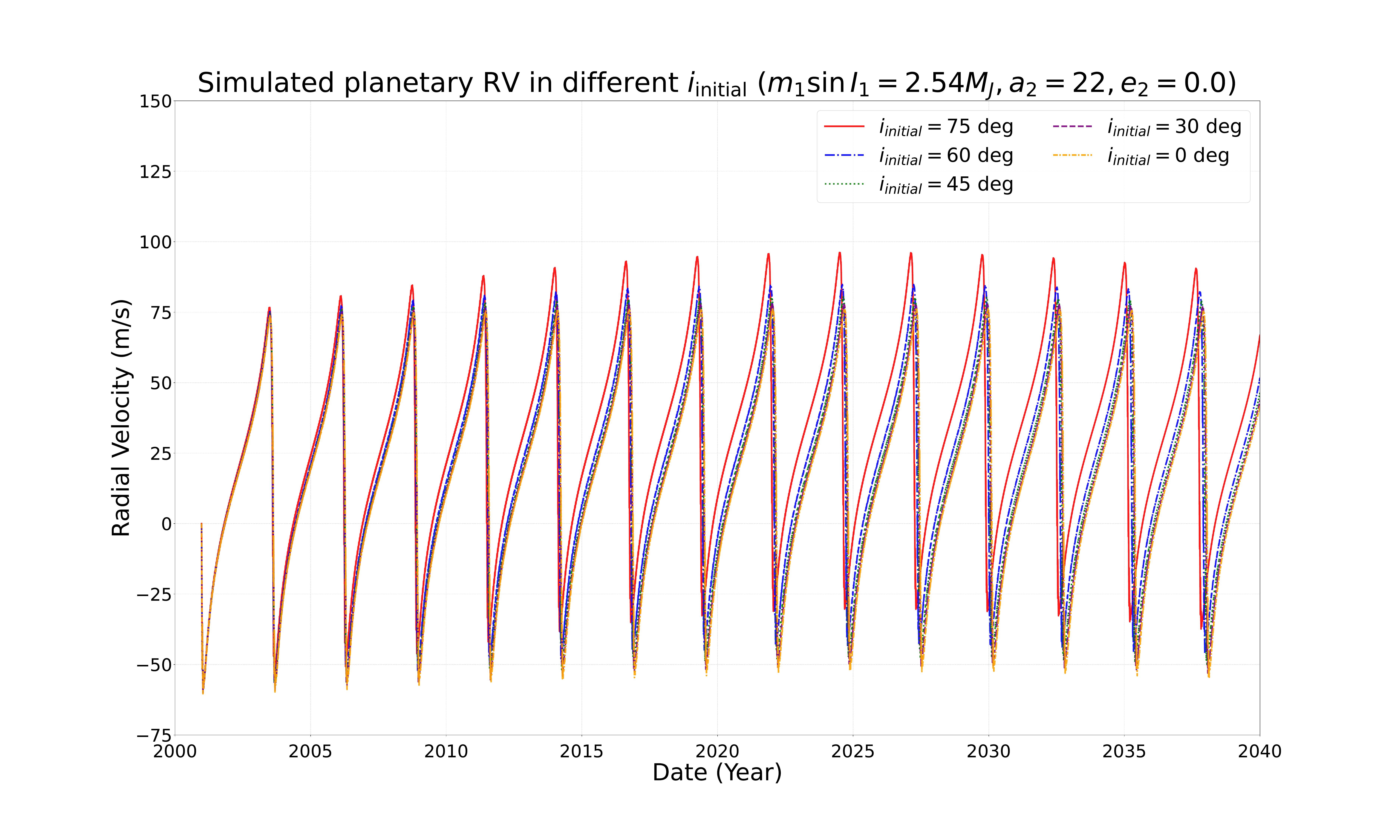}
\includegraphics[width=2\columnwidth]{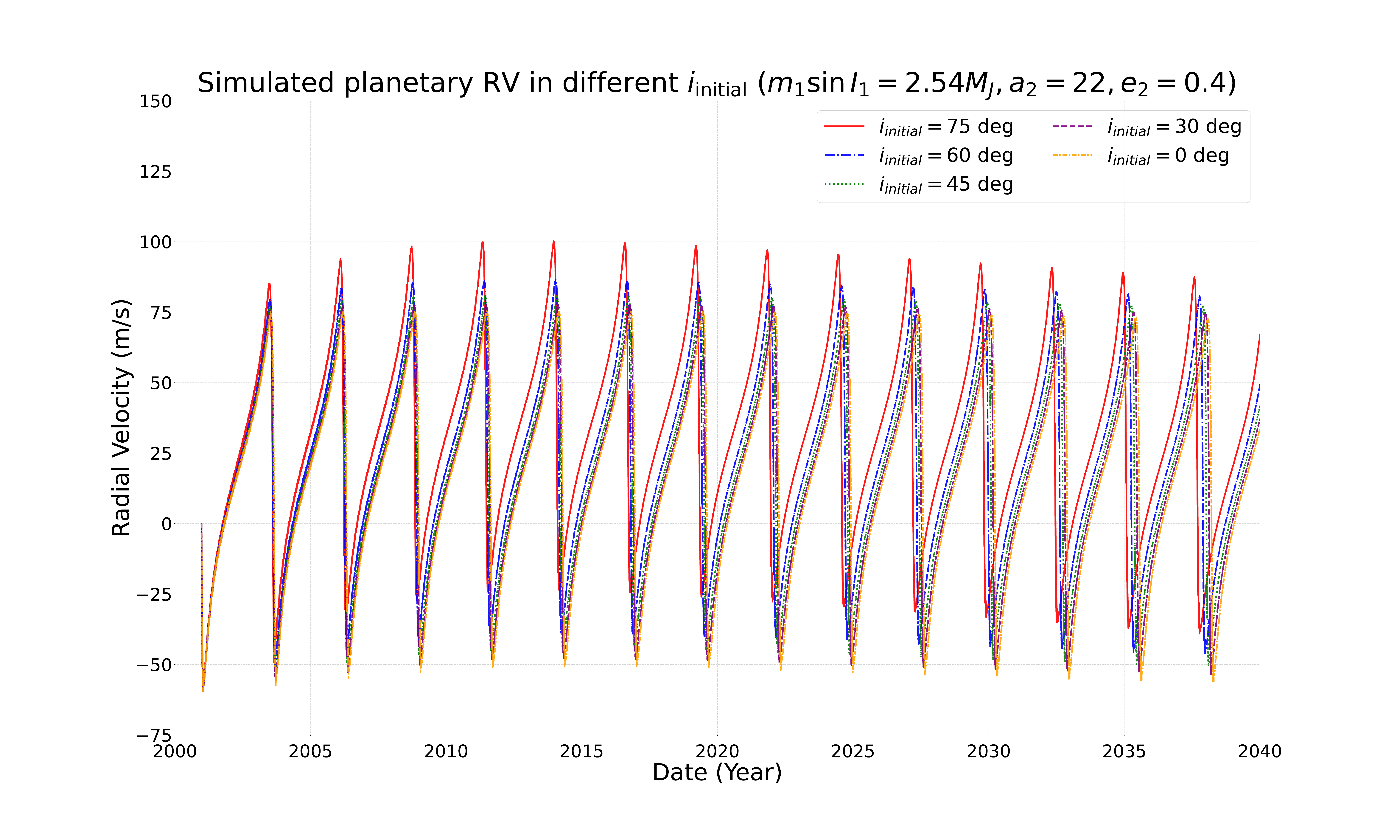}
\caption{
\textbf{Simulated RV curves of HD~41004~Ab for various initial mutual inclinations ($i_{\text{initial}}$) at $a_2 = 22$ AU and $m_1\sin{I_1} = 2.54~M_\textrm{J}$, with $e_2 = 0$ (top panel) and $e_2 = 0.4$ (bottom panel).} 
The RV signals are plotted over a 40-year baseline, illustrating the influence of different initial mutual inclinations ($i_{\text{initial}}$). In the \textbf{top panel}, the configuration with $e_2 = 0$ is shown for $i_{\text{initial}}$ ranging from $0^\circ$ to $75^\circ$. The distinct color curves represent varying inclinations, with higher inclinations displaying larger RV amplitudes. 
The \textbf{bottom panel} presents the corresponding configurations for $e_2 = 0.4$, with the same range of initial inclinations.}
\label{fig:2D}
\end{figure*}

Although a complete vZLK cycle—typically spanning several thousand years—is beyond the reach of the observation, its dynamical imprints can still be detected on much shorter timescales. The two panels of Figure~\ref{fig:2D} show simulated planetary RV curves over a 40-year baseline for different $i_{\text{initial}}$, with a fixed binary separation $a_2 = 22$~AU and a binary eccentricity of $e_2 = 0$ and $e_2 = 0.4$, respectively. At high mutual inclinations, perturbations from the binary induce a systematic drift in the planetary RV signal: $i_{\text{initial}} = 75^\circ$ leads to $\sim 5~\textrm{m s}^{-1}$ per planetary orbit for $e_2 = 0$ and $\sim 10~\textrm{m s}^{-1}$ for $e_2 = 0.4$, while $i_{\text{initial}} = 60^\circ$ shows approximately half of these values. This drift is most prominent near the periastron, where the high orbital velocity amplifies small perturbations. Secular precession of the periastron under vZLK gradually shifts the phase of RV maxima, contributing to the cumulative drift across orbits. In contrast, at lower inclinations, the binary's perturbations are negligible, and the planetary RV curves remain nearly unchanged throughout the 40-year baseline.

In summary, although a full vZLK cycle spans thousands of years, its dynamical effects can manifest on much shorter timescales, particularly in systems with high mutual inclinations. In addition, eccentric vZLK effect can further accelerate these variations. Over decadal baselines, high-inclination configurations exhibit detectable deviations from near-coplanar motion, including secular RV drifts. These short-term signatures of coupled oscillations in inclination and eccentricity are most pronounced near the periastron, which can be captured by current and future high-cadence RV measurements. Systems like HD 41004 therefore serve as ideal laboratories for probing secular evolution via multi-decade RV campaigns, resolving the fundamental mass-inclination degeneracy unattainable with snapshot RV measurements.

\section{Discussion}
\label{sec:discussion}

\subsection{Observational Prospects for Dynamical Signatures}
The activation and amplitude of the vZLK oscillations are governed by mutual inclination $i_{\text{initial}}$, while their period and evolutionary timescales are controlled primarily by binary orbital parameters $a_2$ and $e_2$ (see Section \ref{sec:rv_results} for detailed discussion). In general, smaller $a_2$, higher $e_2$, and larger $i_{\text{initial}}$ lead to shorter modulation timescales and stronger RV drifts. For high-inclination configurations ($i_{\text{initial}} \gtrsim 60^\circ$) in compact binaries ($a_2 \lesssim 40$ AU), vZLK-induced RV deviations become detectable within 10-40 year baselines at $\geq 5\sigma$ confidence with current RV precision ($\sim 1\, \textrm{m s}^{-1}$).

We note that the reported minimum mass of planet Ab carries a significant uncertainty \citep{Zucker_2004}, propagating to the RV semi-amplitude $K_1$. As shown by Equation \ref{RV_equation_general}, the peak $K_1$ at $i_{\text{initial}} = 75^\circ$ increases linearly from $\sim 530\, \textrm{m s}^{-1}$ at $m_1 \sin{I_1} = 1.8~M_\textrm{J}$ to $\sim 950\, \textrm{m s}^{-1}$ at $m_1 \sin{I_1} = 3.28~M_\textrm{J}$, with the reference value $\sim 750\, \textrm{m s}^{-1}$ at $m_1 \sin{I_1} = 2.54~M_\textrm{J}$. Nonetheless, over 40-year baselines, these amplitude variations induce RV drifts below $0.5 \, \text{m s}^{-1}\, \text{yr}^{-1}$ superimposed on the underlying vZLK-driven deviations exceeding $\sim 1.9 \,\text{m s}^{-1}\, \text{yr}^{-1}$ in high-inclination configurations.

Next-generation spectrographs will revolutionize the detection of these dynamical signatures. Instruments like ESPRESSO \citep{Pepe2021} and the upcoming ANDES/ELT \citep{Marconi2021} will achieve long-term RV precision of $\sim10~\textrm{cm s}^{-1}$ to $1~\textrm{m s}^{-1}$. Combined with astrometric constraints—particularly from missions like CHES, capable of delivering $\sim 0.1$AU orbital precision within 100 pc \citep{Feng_2019, Feng_2022, 2022Ji, huang2025closebyhabitableexoplanetsurvey} — these advances enable refined dynamical modeling and detection of sub-$\textrm{m s}^{-1}\textrm{yr}^{-1}$ secular trends within decadal baselines.

\subsection{Impacts of general relativity and tidal effects}

In our dynamical modeling, we have exclusively focused on the vZLK driven evolution, deliberately neglecting short-range effects-particularly general relativistic (GR) precession and tidal interactions. These processes can significantly modify or even suppress vZLK oscillations when their characteristic precession rates become comparable to or faster than the vZLK-induced pericenter precession rate \citep{Fabrycky2007,2015Liu,Naoz_2016, Lu_2025}.

Under the quadrupole approximation, the vZLK-induced pericenter precession rate of the inner orbit is given by:
\begin{align}
\dot{\omega}_{1,\text{vZLK}} = 6C_2 \bigg[& \frac{1}{G_1} ( 4\cos^2 i_{\textrm{initial}} + (5\cos 2\omega_1 - 1)
\notag \\
&\times (1 - e_1^2 - \cos^2 i_{\textrm{initial}}) )\\
&+ \frac{\cos i_{\textrm{initial}}}{G_2} \left( 2 + e_1^2 (3 - 5\cos 2\omega_1) \right) \bigg],\notag
\end{align}
where
\begin{align}
C_2 = \frac{\textrm{G}^4}{16} \frac{(m_0 + m_1)^7}{(m_0 + m_1 + m_2)^3} \frac{m_2^7}{(m_0 m_1)^3} \frac{L_1^4}{L_2^3 G_2^3}.
\end{align}
The pericenter precession rate due to GR is given by \citep{Antoniciello_2021}:
\begin{align}
\dot{\omega}_{1,\text{GR}} = \frac{3 \textrm{G}^{3/2} (m_0 + m_1)^{3/2}}{a_1^{5/2} c^2 (1 - e_1^2)}.
\end{align}
The pericenter precession rate due to tides is given by \citep{Fabrycky2007}:
\begin{align}
\dot{\omega}_{1,\text{Tide}} =& \frac{15 \sqrt{\textrm{G} (m_0+ m_1)}}{8 a_1^{13/2}}\frac{8 + 12e_1^2 + e_1^4}{(1 - e_1^2)^{9/2}} \notag \\
&\times \frac{1}{2} \left[ \frac{m_1}{m_0} k_{0} R_{0}^5 + \frac{m_{0}}{m_1} k_1 R_1^5 \right].
\end{align}
The parameters $k_0$ and $k_1$ are the tidal Love numbers of the star and the planet.

\begin{figure*}[htbp]
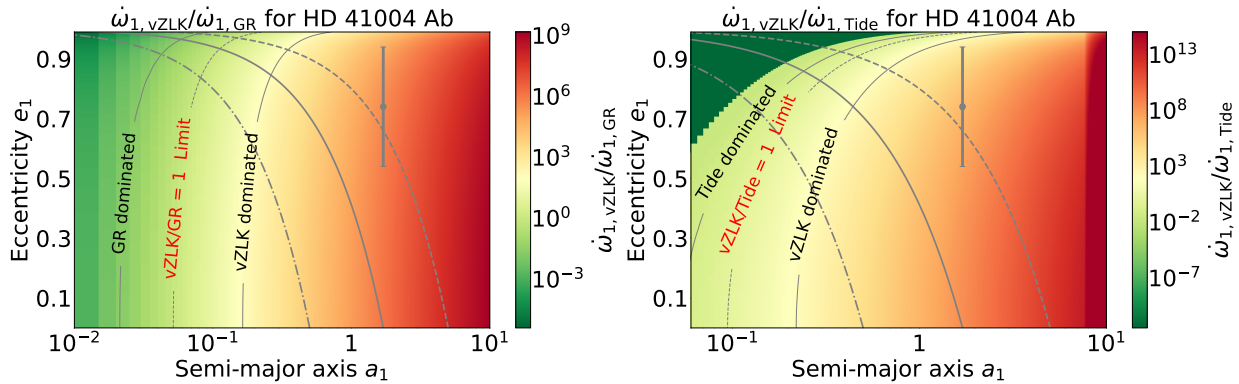

\centering
\includegraphics[width=0.95\columnwidth]{fig6a.pdf}
\includegraphics[width=0.95\columnwidth]{fig6b.pdf}
\caption{
\textbf{Ratio of vZLK to GR-induced pericenter precession rates ($\dot{\omega}_{1,\text{vZLK}} / \dot{\omega}_{1,\text{GR}}$) and tide-induced pericenter precession rates ($\dot{\omega}_{1,\text{vZLK}} / \dot{\omega}_{1,\text{Tide}}$) for an HD~41004~Ab-like planet as a function of semimajor axis ($a_1$) and eccentricity ($e_1$).}  
The color map illustrates the relative dominance of general relativistic and tidal (GR, Tide; green) versus vZLK-induced (red) precession across the $(a_1, e_1)$ parameter space. The gray point with error bars marks the nominal location of HD~41004~Ab, with $a_1 = 1.7\,\mathrm{AU}$ and $e_1 = 0.74 \pm 0.2$. Three gray contours denote the periastron distances typical of warm Jupiters ($a_1=$ 0.5 AU, dash-dotted line), HD~41004~Ab itself (solid line), and cold Jupiters ($a_1=$ 5 AU, dashed line). The system resides well within the vZLK-dominated regime, with a precession rate ratio $\dot{\omega}_{1,\text{vZLK}} / \dot{\omega}_{1,\text{GR}}$ of $\sim 3.5 \times 10^5$  and $\dot{\omega}_{1,\text{vZLK}} / \dot{\omega}_{1,\text{Tide}}$ of $\sim1.8\times 10^{13}$ at $i_{\text{initial}} = 75^\circ$.
}
\label{fig:ratio}
\end{figure*}

We compute the precession rate ratios $\dot{\omega}_{1,\text{vZLK}} / \dot{\omega}_{1,\text{GR}}$ and $\dot{\omega}_{1,\text{vZLK}} / \dot{\omega}_{1,\text{Tide}}$ for an HD 41004 Ab-like planet across a range of semi-major axes and eccentricities at $m_1\sin{I_1}=2.54M_\textrm{J}$. We adopt typical Love numbers of $k_0 = 0.025$ for the primary star and $k_1 = 0.3$ for the planetary companion, representative of solar-type stars and giant planets \citep{2014Ogilvie}. The stellar and planetary radii are set to $R_0 = 0.7\,R_{\odot}$ and $R_1 = 1\,R_{\mathrm{J}}$, respectively. The results are shown in Figure~\ref{fig:ratio}.

The precession rate ratio $\dot{\omega}_{1,\text{vZLK}} / \dot{\omega}_{1,\text{GR}}$ reaches approximately $3.5 \times 10^5$, while $\dot{\omega}_{1,\text{vZLK}} / \dot{\omega}_{1,\text{Tide}}$ is around $1.8 \times 10^{13}$ at the nominal system parameters ($a_1 = 1.7$~AU, $e_1 = 0.74$). Taking the simulated vZLK-induced RV drift (\(\sim 1.9~\mathrm{m\,s^{-1}\,yr^{-1}}\) at \(i_{\text{initial}} = 75^\circ\)) as a reference, and scaling it using the precession rate ratios and the Keplerian dependence on \(\omega_1\), we estimate 10-year RV variations of only \(\sim 5.4 \times 10^{-5}~\mathrm{m\,s^{-1}}\) for GR and \(\sim 1.1 \times 10^{-12}~\mathrm{m\,s^{-1}}\) for tidal precession. These results indicate that the vZLK mechanism overwhelmingly dominates the precession dynamics. Similar outcomes are found at other inclinations, highlighting that GR and tidal precession are insignificant for HD 41004 Ab at present.

For short-period S-type planets in other compact binaries, e.g. GJ~86 and $\tau$~Bootis, short-range forces may significantly affect the dynamics and should be taken into account to describe their dynamical evolutions accurately.

%Building on our results for HD~41004, we plan to extend our vZLK-driven modeling framework to other compact binary systems. \textbf{Future high-precision RV and astrometric observations will also be essential for refining system parameters and validating dynamical predictions.} Together, these efforts aim to enhance our understanding of planet formation and evolution in close binary environments.

\section{Conclusions}
\label{sec:conclusion}

Through dynamical imprints on observables, secular interactions and mean-motion resonances (MMRs) in multi-planet systems constrain orbital architectures—as demonstrated by \citet{Correia_2010} for mutual inclinations, \citet{Judkovsky_2022} for long-term trends, and \citet{huang2025closebyhabitableexoplanetsurvey} for MMR-induced RV signatures. 

For S-type planets in compact binaries, secular perturbations may be dominated by the vZLK mechanism—replacing MMRs with binary-driven oscillations as the primary dynamical architect. In this study, we investigate the dynamical stability and potential observational signatures of vZLK cycles in the HD~41004 system. Our main conclusions are as follows:

\begin{itemize}
\item The planet HD~41004~Ab can remain dynamically stable with mutual inclinations ($i_\textrm{initial}$) up to approximately $75^\circ$ across various system configurations. The stability analysis has taken into account the fact that the true mass of the planet varies with angle $I_1$ because of a fixed minimum mass $m_1\sin I_1$ obtained by RV measurements. Three-body dynamical RV fitting for HD~41004~A across $a_2$ ranging from 22 to 40 AU and low to moderate $e_2$ values demonstrates comparable accuracy to Keplerian fits for different $i_\textrm{initial}$. Both stability simulations and RV modeling do not rule out the possibility of a close binary configuration with low or moderate eccentricities. This result highlights that orbits of S-type planets in compact binary systems can maintain stable even under strong secular perturbations.

\item Above a critical mutual inclination angle, the vZLK mechanism causes long-term variations in eccentricity, inclination, and apsidal orientation of the planet. With unrestricted three-body simulations, we show that the vZLK timescale shortens significantly with increasing mutual inclination, while the amplitude of inclination oscillations also becomes larger. As a result, highly inclined configurations exhibit stronger and faster RV variations, enhancing the chance of directly observing vZLK effects on accessible timescales. The presence of eccentric vZLK can accelerate these drifts, enhancing the detectability.

\item Future RV measurements near periastron could differentiate between high and low mutual inclinations that are both dynamically permitted in this system. At high inclinations, the effects are particularly pronounced, with RV drifts reaching over $\sim 5~\textrm{m s}^{-1}$ per  planetary orbit ($\sim 1.9 \,\text{m s}^{-1}\, \text{yr}^{-1}$) at $i_{\text{initial}} = 75^\circ$ and $\sim 2.5~\textrm{m s}^{-1}$ per planetary orbit ($\sim 0.9 \,\text{m s}^{-1}\, \text{yr}^{-1}$) at $i_{\text{initial}} = 60^\circ$ in circular binary configuration. In contrast, low-inclination configurations exhibit minimal secular modulation over similar timescales, requiring extended baselines or complementary astrometric data to constrain system parameters. 
Nonetheless, determining the planet's true mass can benefit from constraints on the mutual inclination.
\end{itemize}

Our results suggest that HD~41004 and other similar S-type planetary systems, e.g. GJ~86 \citep{2000Queloz,2022Zeng}, $\tau$~Bootis \citep{1997Butler,2019Justesen}, GJ~3021 \citep{2001Naef}, HD~196885 \citep{Chauvin2011,2023Chauvin}, and HD~164509 \citep{2012Giguere}, are promising targets of studying consequences of secular dynamical effects. Future high-precision RV observations, particularly when combined with astrometric measurements, will enable tighter constraints on the architecture of those hierarchical systems such as mutual orbital inclination, planetary mass, as well as long-term dynamical evolution.

\section*{Acknowledgement}
We thank the referee for many insightful comments that greatly improved the quality of this paper. We also thank Zhecheng Hu, Xiumin Huang, Jianghui Ji, Man Hoi Lee, Doug Lin, Bin Liu, Chris Ormel, Xuesong Wang, and Zixin Zhang for insightful discussions. S.-F.L. acknowledges the support from the Guangdong Basic and Applied Basic Research Foundation under grant No. 2021B1515020090,  the National Natural Science Foundation of China under grant No. 11903089, and the China Manned Space Program under grant No. CMS-CSST-2025-A16. B.M. acknowledges the ET2.0 project funding from Shanghai Astronomical Observatory. Z.Q. would like to thank the hospitality of the Proto-planetary Disk and Planet Formation Summer School
organized by Xuening Bai and Ruobing Dong in 2022,
hosted by the Chinese Center for Advanced Science and
Technology.

%\vspace{5mm}
%\facilities{HST(STIS), Swift(XRT and UVOT), AAVSO, CTIO:1.3m,


%\software{astropy \citep{2013A&A...558A..33A,2018AJ....156..123A},  
%Cloudy \citep{2013RMxAA..49..137F}, 
%Source Extractor \citep{1996A&AS..117..393B}
%}



\appendix
\section{MEGNO Stability Analysis}
\label{sec:megno}

The Mean Exponential Growth factor of Nearby Orbits (MEGNO)\citep{megno} is a widely used chaos indicator that efficiently distinguishes between stable, quasiperiodic trajectories and chaotic, unstable motion, while also highlighting underlying resonance structures \citep{Maffione_2011}. This makes it particularly well-suited for hierarchical three-body systems such as HD~41004, where secular perturbations and strong dynamical coupling can drive complex, long-term orbital evolution.

The MEGNO parameter, $Y(\phi(t))$, quantifies the mean exponential divergence rate between a reference orbit $\phi(t)$ and a nearby orbit $\phi'(t)$ that differs by an infinitesimal perturbation in initial conditions. It is formally defined as:
\begin{equation}
Y(\phi(t)) = \frac{2}{t}\int^t_0 \frac{\dot{\delta}(\phi(s))}{\delta(\phi(s))}sds,
\label{megno}
\end{equation}
where $\delta(\phi(t))$ represents the time-dependent separation between the two orbits, and $\dot{\delta}(\phi(t))$ is its time derivative. The integrand, $\dot{\delta} / \delta$, corresponds to the instantaneous logarithmic divergence rate, directly related to the maximum Lyapunov Characteristic Exponent (mLCE).

The time-averaged MEGNO value, $\bar{Y}(\phi(t))$, is given by:
\begin{equation}
\bar{Y}(\phi(t)) = \frac{1}{t}\int^t_0Y(\phi(s))ds.
\end{equation}
For stable, quasiperiodic orbits, $\bar{Y}(\phi(t))$ converges asymptotically to 2 as $t \to \infty$:
\begin{equation}
\bar{Y}(\phi(t)) \simeq 2 - \frac{2 \ln(1+\lambda t)}{\lambda t} + O(\phi(t)) \to 2,
\end{equation}
where $\lambda$ is the mLCE. Conversely, for chaotic or unstable orbits, $\bar{Y}(\phi(t))$ diverges from 2 with increasing integration time, clearly signaling dynamical instability.

In the HD~41004 system, instability often manifests through the vZLK-driven excitation of large eccentricities, which may bring the planet's periastron $a_1(1-e_1)$ close to the stellar Roche limit, leading to tidal disruption or ejection. In our stability assessment, orbits are classified as unstable if the MEGNO value fails to converge to 2 or if the periastron distance falls below the Roche limit during the simulation.

\section{Dynamical Fitting of Existing RV data for Different Parameters}
\label{sec:rv_fitting_2}

\begin{figure*}[htbp]
\centering
\includegraphics[width=1\columnwidth]{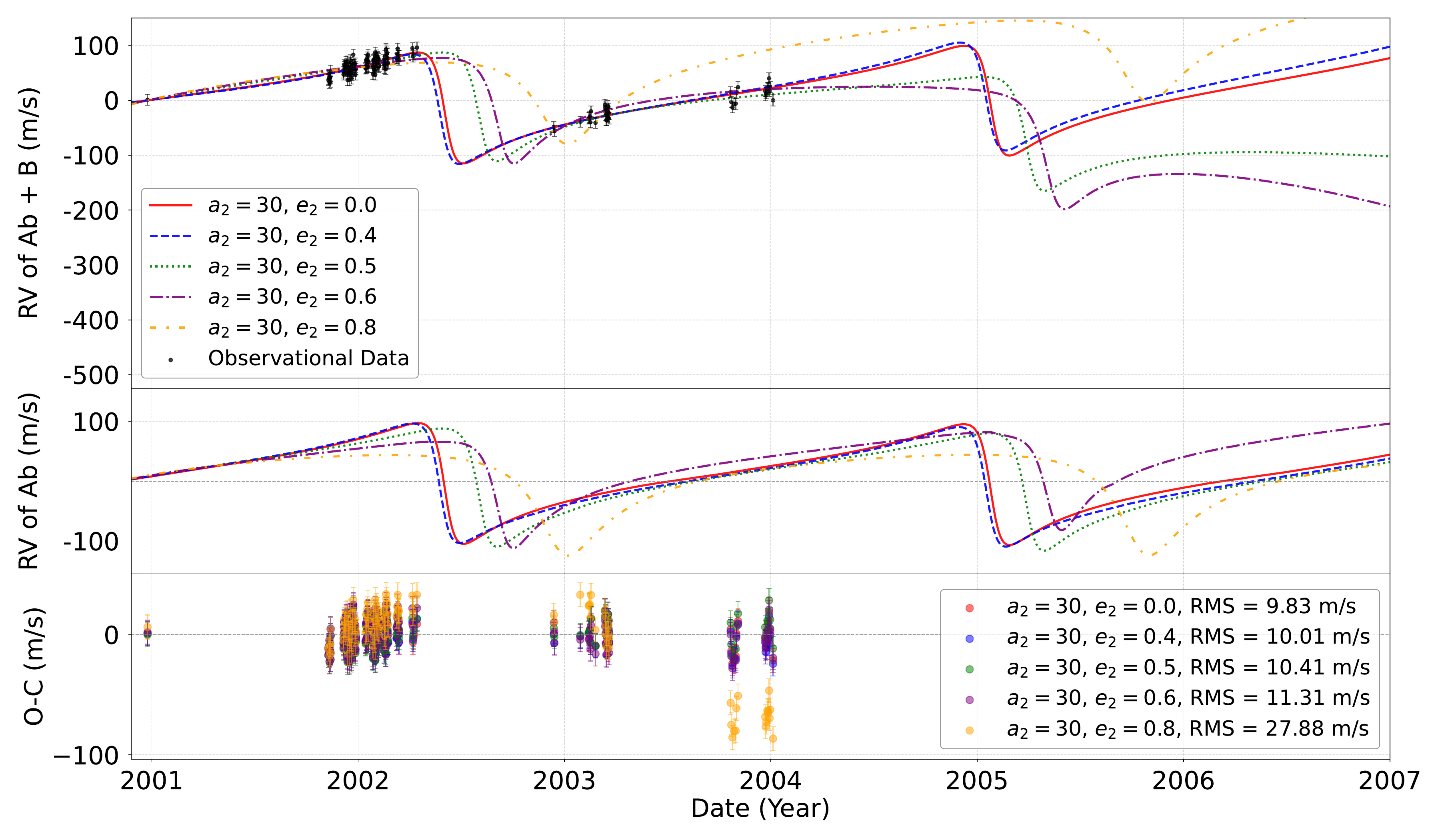}
\includegraphics[width=1\columnwidth]{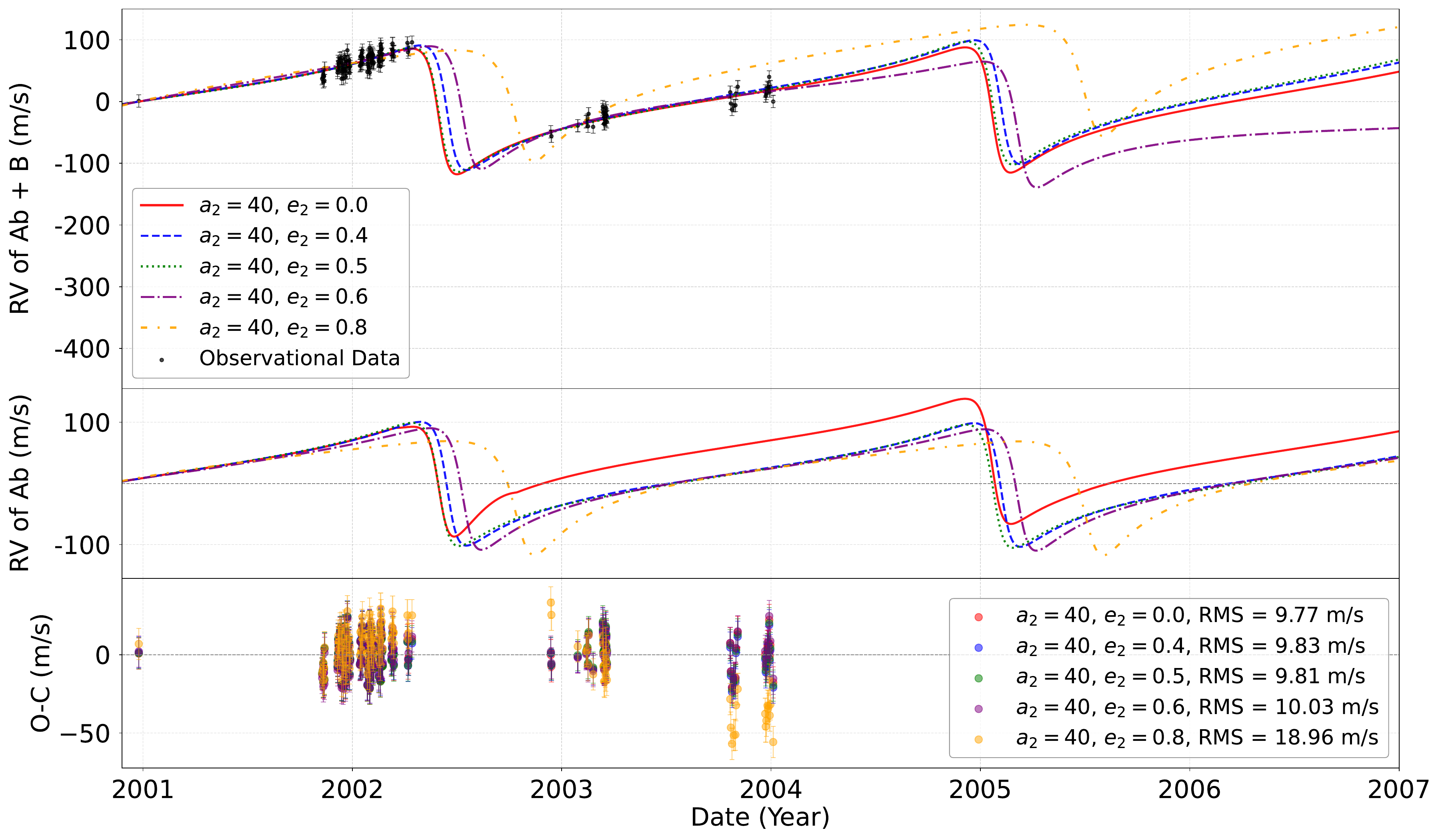}
\caption{
\textbf{Dynamical RV Fitting of HD~41004~A for $a_2=30$ AU and $a_2=40$ AU.} The structure is the same as Figure~\ref{fig:RV_fit}. Different colors and line styles indicate varying eccentricities. The planetary mass is fixed at $m_1\sin{I_1} = 2.54~M_\textrm{J}$. RMS values of the O-C residuals are listed in the legends.
}
\label{fig:rv_fit_appendix_a2}
\end{figure*}

\begin{figure*}[htbp]
\centering
\includegraphics[width=1\columnwidth]{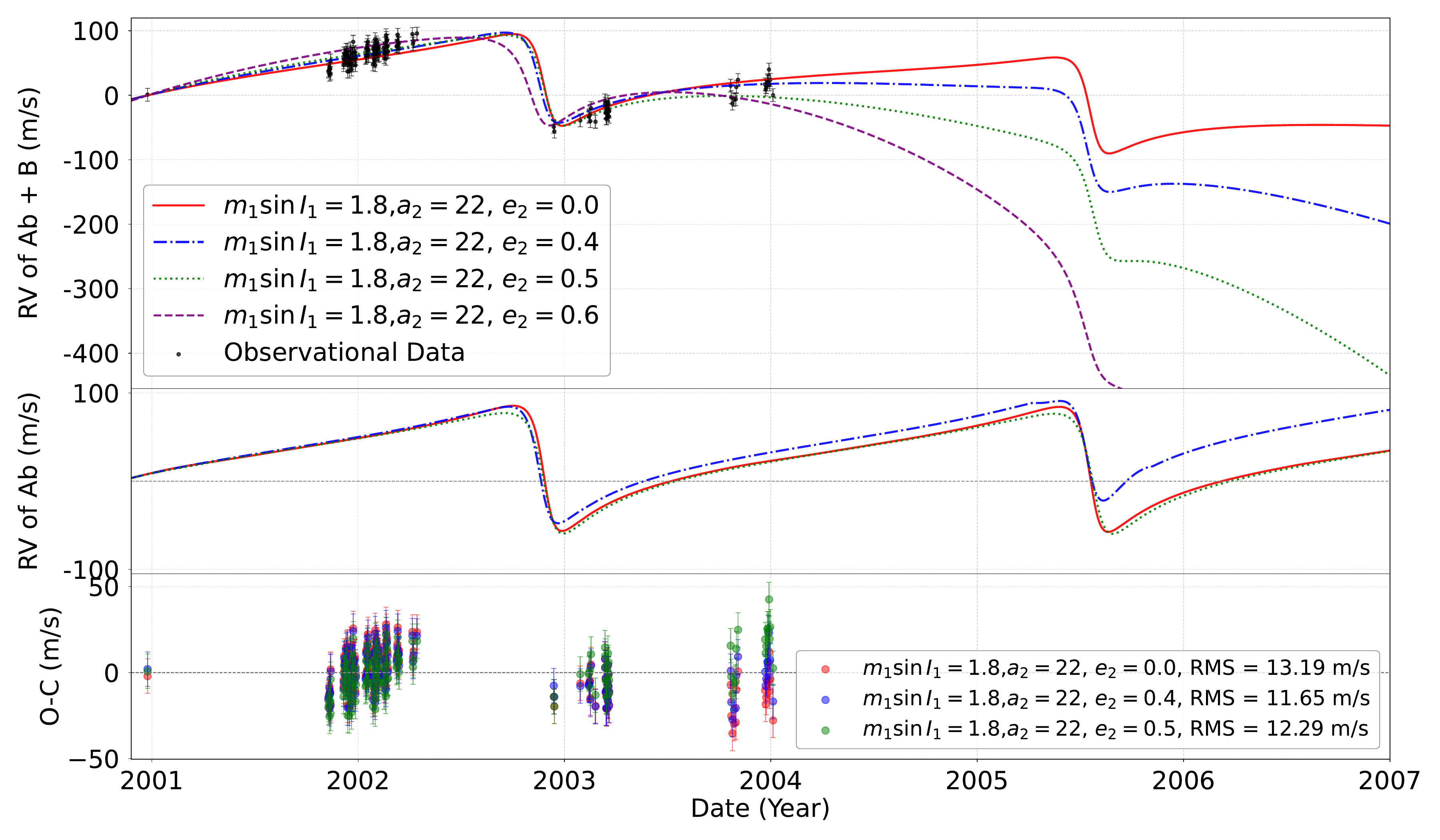}
\includegraphics[width=1\columnwidth]{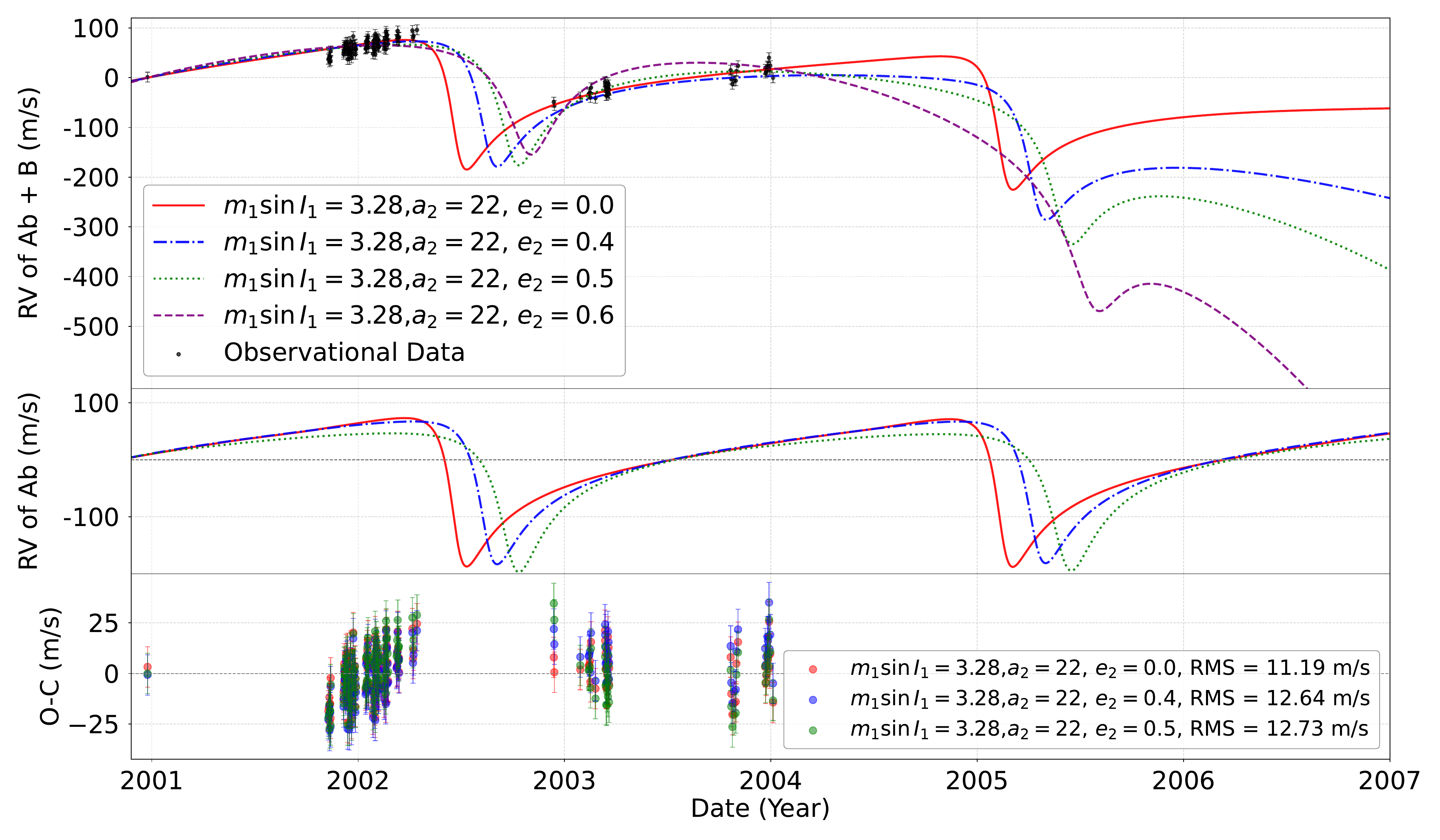}
\caption{
\textbf{Dynamical RV Fitting of HD~41004~A for different planetary masses at $a_2 = 22$ AU.} 
The structure is the same as Figure~\ref{fig:RV_fit}. The \textbf{top panel} represents fits for $m_{\text{pl}} = 1.8~M_\textrm{J}$, while the \textbf{bottom panel} corresponds to $m_{\text{pl}} = 3.28~M_\textrm{J}$.}
\label{fig:rv_fit_appendix_mass}
\end{figure*}

Figures \ref{fig:rv_fit_appendix_a2} and \ref{fig:rv_fit_appendix_mass} present additional RV modeling of HD~41004~A under different configurations, complementing the analysis in Section~\ref{subsec:rv_fit}. The structure of these figures follows the same format as Figure~\ref{fig:RV_fit}, with the top, middle, and bottom panels displaying the combined RV signal, the isolated planetary signal, and the O-C residuals, respectively.

Figure~\ref{fig:rv_fit_appendix_a2} investigates the impact of different semi-major axes ($a_2 = 30$ AU and $40$ AU) for a fixed minimum planetary mass of $m_1\sin{I_1} = 2.54~M_\textrm{J}$. Figure~\ref{fig:rv_fit_appendix_mass} explores the effect of varying minimum planetary masses ($m_1\sin{I_1} = 1.8~M_\textrm{J}$ and $3.28~M_\textrm{J}$) at $a_2 = 22$ AU. The eccentricities of the stellar companion ($e_2$) are varied in both cases to assess their influence on the RV curves.

\section{Timescale of vZLK cycles}
\label{sec:vZLK_timescale}
We conduct unrestricted three-body simulations based on the dynamical RV model over a grid of initial mutual inclinations ranging from $0^\circ$ to $80^\circ$, measuring the vZLK timescale for each configuration with year-level precision. For each run, we also record the amplitude of inclination oscillations ($i_{\min}$ to $i_{\max}$) to quantify the extent of vZLK-induced variations. 

Given the uncertainty in $e_2$ and the instability observed at smaller $a_2$ for $e_2 = 0.8$, together with the RV fitting results in Section~\ref{subsec:rv_fit} that favor low to moderate binary eccentricities, our simulations focus on $e_2 = 0$ and $e_2 = 0.4$ at $a_2 = 22$~AU with $m_1\sin{I_1}=2.54\,M_\textrm{J}$. Although wider $a_2$ values are still dynamically viable, they exhibit similar secular behavior and are sufficiently captured by analogous cases.
\begin{figure}[htbp]
\centering
\begin{minipage}{0.48\textwidth}
    \centering
    \includegraphics[width=\textwidth]{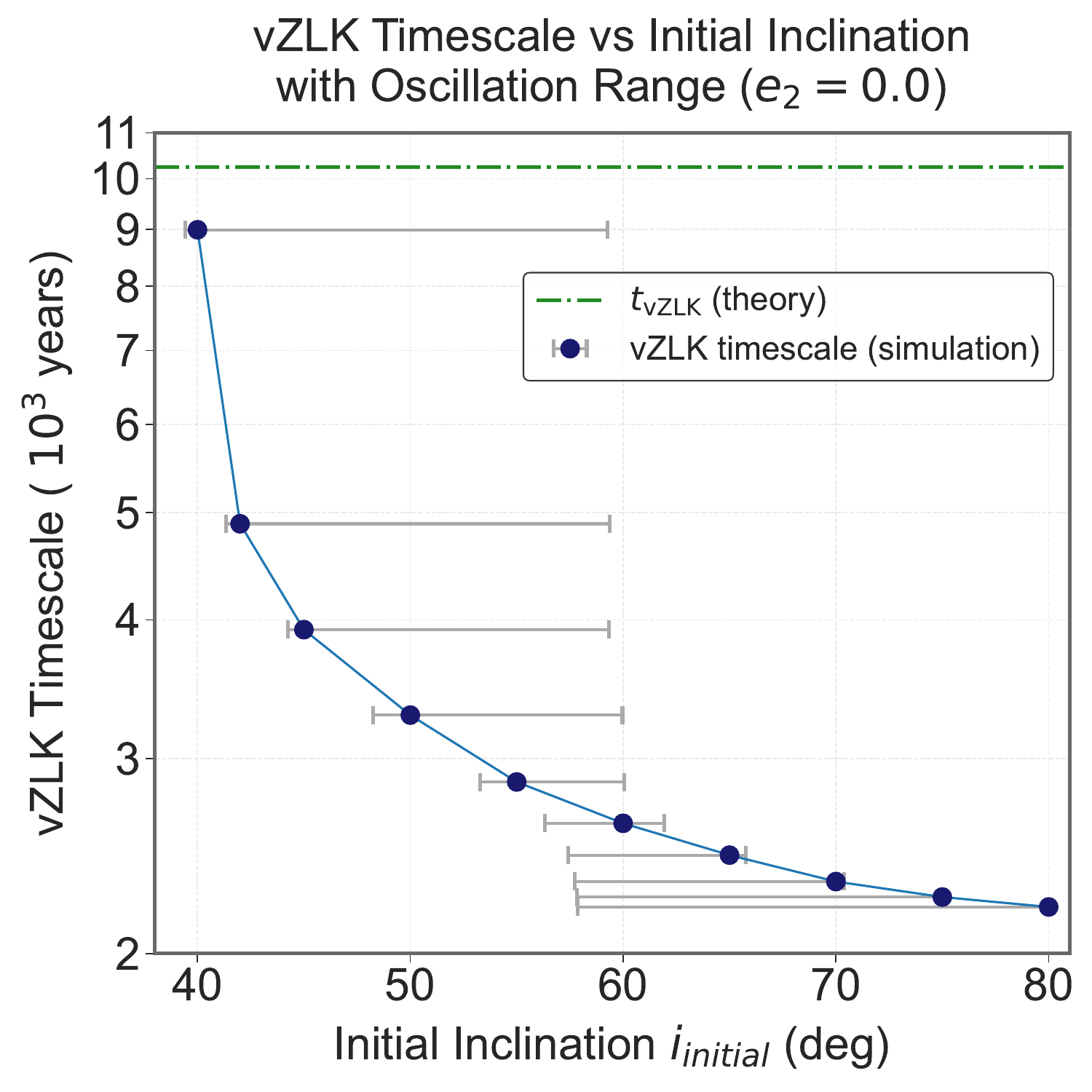}
\end{minipage}
\hfill
\begin{minipage}{0.48\textwidth}
    \centering
    \includegraphics[width=\textwidth]{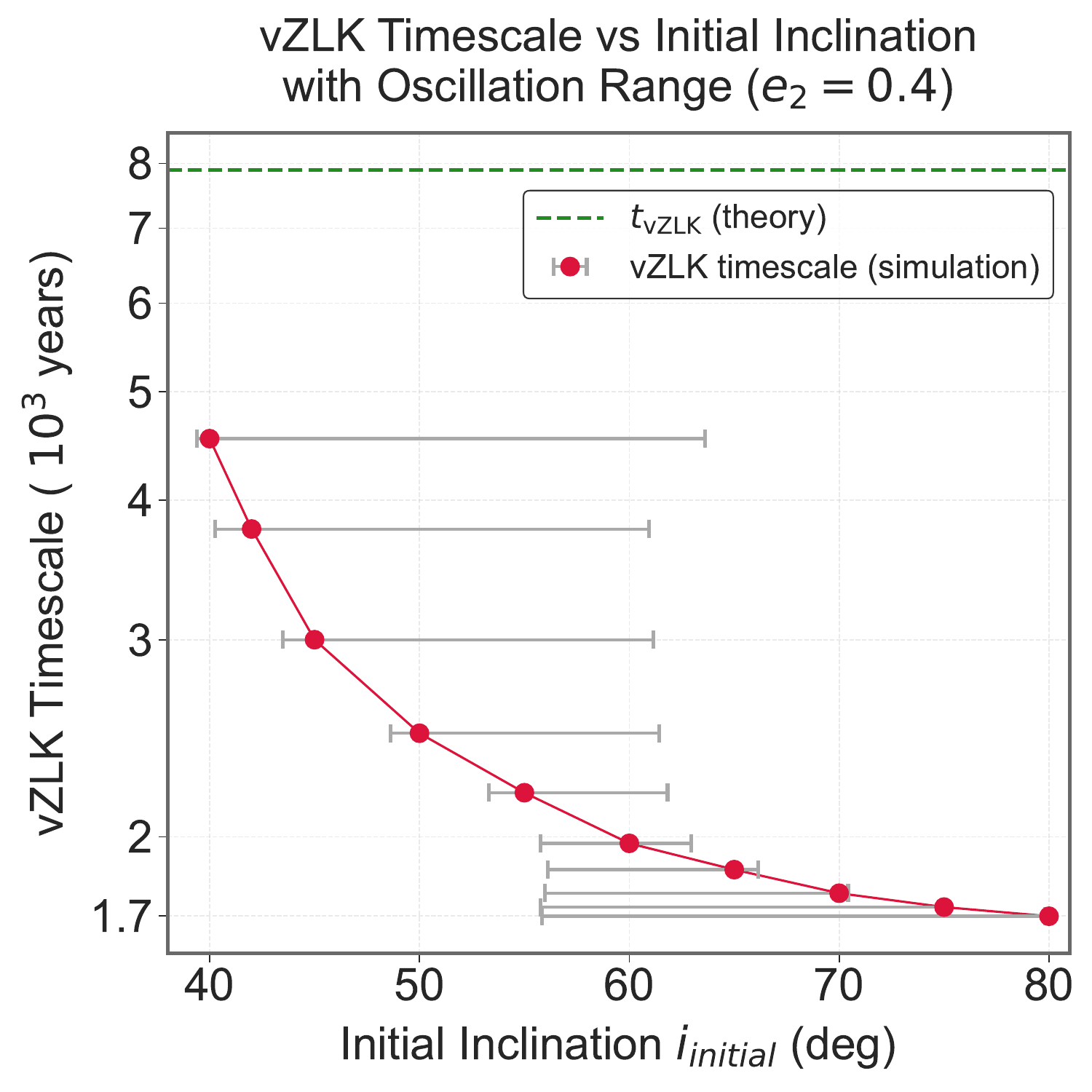}
\end{minipage}
\caption{
\textbf{Dependence of the vZLK timescale on the initial mutual inclination in the HD~41004 system for different binary eccentricities.}
Solid points indicate the simulated vZLK timescales as functions of $i_{\text{initial}}$, with horizontal error bars denoting the corresponding inclination oscillation amplitudes (from $i_{\min}$ to $i_{\max}$). The green dash-dotted lines mark the theoretical vZLK timescales derived from the test-particle approximation, calculated as $1.02 \times 10^4$ years for $e_2 = 0$ and $7.89 \times 10^3$ years for $e_2 = 0.4$. Both panels reveal a systematic decrease in the vZLK timescale with increasing inclination, highlighting enhanced dynamical interactions at higher $i_{\text{initial}}$ values.
}
\label{fig:vZLK_timescale}
\end{figure}

Figure~\ref{fig:vZLK_timescale} shows the simulation results. In the left panel ($e_2 = 0$), the simulated vZLK timescale systematically decreases with increasing mutual inclination, from $\sim 8.99 \times 10^3$ years at $i_{\text{initial}} = 40^\circ$ to $\sim 2.20 \times 10^3$ years at $80^\circ$. The right panel ($e_2 = 0.4$) shows a similar trend, with timescales decreasing from $\sim 4.53 \times 10^3$ years to $\sim 1.70 \times 10^3$ years over the same inclination range. These values remain consistently below the theoretical prediction, also underscore the amplifying effect of binary eccentricity.

The error bars in both panels indicate the oscillation ranges of inclination during the vZLK cycles, with larger spreads at high inclinations. This behavior demonstrates that high-inclination orbits experience both faster vZLK oscillations and larger variations, enhancing the potential for detectable long-term RV signals.

\bibliography{ref}{}
\bibliographystyle{aasjournal}

\end{CJK*}
\end{document}